\documentclass[11pt,a4paper]{article}
\pdfoutput=1
\usepackage{jheppubedit}
\usepackage{enumitem}
\usepackage{amsmath}
\usepackage{amsfonts}
\usepackage{amssymb}
\usepackage{mathtools}
\usepackage{physics}
\usepackage{graphicx}
\usepackage{caption}
\usepackage{tensor}
\usepackage{tikz}
\usepackage{simplewick}

\newcommand{\dpi}{\mathcal{D}}
\newcommand{\mj}{\mathcal{J}}

\newcommand{\mq}{\mathcal{Q}}
\def\e{\mbox{{\rm e}}}
\newcommand{\tj}[6]{ \begin{pmatrix}
   #1 & #2 & #3 \\
   #4 & #5 & #6 
  \end{pmatrix}}
  \newcommand{\sj}[6]{ \begin{Bmatrix}
   #1 & #2 & #3 \\
   #4 & #5 & #6 
  \end{Bmatrix}}

\title{\center{Edge Dynamics from the Path Integral\\\Large{Maxwell and Yang-Mills}}}

\author[a]{Andreas Blommaert}
\author[a]{,Thomas G. Mertens}
\author[a]{and Henri Verschelde}
\affiliation[a]{Ghent University, Department of Physics and Astronomy\\
Krijgslaan, 281-S9, 9000 Gent, Belgium.}

\emailAdd{andreas.blommaert@ugent.be}
\emailAdd{thomas.mertens@ugent.be}
\emailAdd{henri.verschelde@ugent.be}
\abstract{We derive an action describing edge dynamics on interfaces for gauge theories (Maxwell and Yang-Mills) using the path integral. The canonical structure of the edge theory is deduced and the thermal partition function calculated. We test the edge action in several applications. For Maxwell in Rindler space, we recover earlier results, now embedded in a dynamical canonical framework. A second application is 2d Yang-Mills theory where the edge action becomes just the particle-on-a-group action. Correlators of boundary-anchored Wilson lines in 2d Yang-Mills are matched with, and identified as correlators of bilocal operators in the particle-on-a-group edge model.}

\begin{document}

\maketitle
\section{Introduction}
Gauge theories with interfaces exhibit degrees of freedom living exclusively on the boundary, the so-called \emph{edge states}. While this phenomenon has been known for quite a while in topological theories (e.g. 3d Chern-Simons theories exhibit a chiral 2d WZW conformal theory on their boundary), it is only relatively recent that this has been studied for gauge theories with propagating degrees of freedom. Edge states have been extensively studied recently in e.g. \cite{Buividovich:2008gq,Casini:2013rba,Casini:2014aia,Radicevic:2014kqa,Radicevic:2015sza,Radicevic:2016tlt,Donnelly:2014gva,Donnelly:2011hn,Donnelly:2012st,Donnelly:2014fua,Donnelly:2015hxa,Huang:2014pfa,Ghosh:2015iwa,Hung:2015fla,Aoki:2015bsa,Soni:2015yga,VanAcoleyen:2015ccp,Casini:2015dsg,Soni:2016ogt,Michel:2016qge,Seraj:2017rzw}.

This field of study is intimately linked with the mysterious contact contribution to the thermal entropy of gauge fields in Rindler space, found by Kabat in \cite{Kabat:1995eq,Kabat:1995jq,Kabat:2012ns} using replica trick methods. Donnelly and Wall found an explanation for this term as a Euclidean path integral over static horizon radial electric fields $E(\mathbf{x})$ weighted by the on-shell Euclidean action \cite{Donnelly:2014fua,Donnelly:2015hxa}: 
\begin{equation}
\label{WDpf}
Z^{\text{edge}}=\int \left[\mathcal{D}E(\mathbf{x})\right]e^{-S\left[E(\mathbf{x})\right]\rvert_\text{on-shell}} \equiv \text{`` Tr} e^{-\beta H}\text{ ''},
\end{equation}
which should be read as implementing the thermal trace, with $S\rvert_\text{on-shell} = \beta H$ and the trace running over all static configurations $E(\mathbf{x})$. This procedure identifies the origin of the contact term as counting electrostatic configurations (or surface charges) on the horizon, but does not reveal the underlying canonical structure associated with these electrostatic boundary degrees of freedom. In a different work \cite{Donnelly:2016auv}, Donnelly and Freidel used the requirement of (large) gauge invariance to write down a presymplectic potential for Maxwell degrees of freedom on a generic boundary: a new scalar field $\phi$ compensates for the large gauge transformation, and becomes a new dynamical field conjugate to the surface charges.

Recently in \cite{Blommaert:2018rsf}, using canonical quantization of the Maxwell action, we presented an alternative procedure that gives rise to a canonical structure on the boundary of Maxwell: instead of saving large gauge-invariance by introducing a new variable, we abandon it completely and promote the large gauge degrees of freedom $\phi$ themselves to physical variables, conjugate to the boundary charges $E(\mathbf{x})$. This promotion of large gauge parameters to physical boundary variables is a key ingredient in topological theories such as 3d Chern-Simons, where they make up the boundary theory dual to the entire bulk theory.
\\~\\
The question arises whether there is a boundary action that gives rise to this canonical structure.\footnote{A similar goal is being pursued by Geiller and Jai-akson (private communication).} Given such an action, it would be possible to study boundary correlation functions and real-time dynamics. In this work, we derive this boundary edge action for Maxwell and Yang-Mills theory. The result is a Lorentzian action that depends on the boundary gauge field $g$ and the boundary current density $\mj$:
\begin{equation}
S^\text{bdy}\left[\mj,g\right]=\int d^{d-1}x\, \Tr(\frac{1}{2}\mathcal{\mj}^\alpha g A[\mj]_\alpha g^{-1} - \mj^\alpha\partial_\alpha g g^{-1}),\label{action}
\end{equation}
where $A[\mj]$ denotes the on-shell evaluation of the gauge field sourced by the boundary current $\mj$.

This action is put to the test in several applications. After briefly discussing Maxwell edge states in flat space, we study the Maxwell edge action in Rindler space where we recover the results of \cite{Donnelly:2014fua,Donnelly:2015hxa,Blommaert:2018rsf}, now placed in a dynamical canonical context. One particularly interesting aspect of this calculation is that it incorporates a natural proof for the absence of horizon degrees of freedom associated with the spatial currents.

Having an action allows us to describe dynamics of edge modes and to compute various correlation functions. We will do this explicitly for the example of 2d Yang-Mills theory on a disk. 2d Yang-Mills has been studied extensively in the past (see e.g. \cite{Migdal:1975zg,Witten:1991we,Witten:1992xu} for some of the foundational results and \cite{Cordes:1994fc} for a particularly nice review and references therein). The boundary edge theory of 2d Yang-Mills on a disk is found to be the particle-on-a-group model. Next to the partition function, also the correlators of this model are known \cite{Mertens:2018fds}. We identify the bulk duals of generic boundary correlators as boundary-anchored Wilson lines. This effectively solves the boundary edge theory of 2d Yang-Mills: in principle we are able to study an arbitrary dynamical process in the boundary.
\\~\\
The paper is organized as follows. In section \ref{sect:Maxwell} we consider Maxwell theory before generalizing in section \ref{sect:YM} to non-abelian Yang-Mills theories. Applications to flat space (section \ref{s:warmup}), Rindler space (sections \ref{s:rindler} and \ref{s:YMrindler}), 2d Maxwell (section \ref{s:2dmw}) and 2d Yang-Mills (section \ref{s:2dym}) are discussed in the process. Correlators of the boundary edge theory of 2d Yang-Mills are discussed in \ref{s:cor}. \\
Throughout this work, indices $\alpha,\beta,\hdots$ live on the boundary $\partial\mathcal{M}$ and indices $\mu,\nu,\hdots$ live on $\mathcal{M}$.

\section{Maxwell}
\label{sect:Maxwell}
In this section, we present the derivation of the boundary action \eqref{action} for Maxwell theory, and put it to the test in two applications: Maxwell in Rindler, and 2d Maxwell which is a quasi-topological theory. 

\subsection{Boundary Action for Maxwell}
\label{s:bdy}
Consider Maxwell theory on the $d$-dimensional Lorentzian manifold $\mathcal{M}\cup \bar{\mathcal{M}}$ without boundaries:
\begin{equation}
S=-\frac{1}{4}\int_{\mathcal{M}\cup \bar{\mathcal{M}}}F\wedge \star F=-\frac{1}{4}\int _{\mathcal{M}\cup \bar{\mathcal{M}}}d^d x\sqrt{-g}\,F^{\mu\nu}F_{\mu\nu}.
\end{equation}
Defining separate fields $A$ and $\bar{A}$, restricted to respectively $\mathcal{M}$ and $\bar{\mathcal{M}}$, we can split the full path integral into two pieces, glued together along the boundary $\partial \mathcal{M}$ using a functional delta-constraint:
\begin{equation}
\int \left[\mathcal{D} A_\mu\right]\left[\mathcal{D} \bar{A}_\mu \right] \prod_{x \in \partial \mathcal{M}} \delta\left(A_{\mu}-\bar{A}_{\mu}\right) e^{i S_A} e^{i S_{\bar{A}}}.
\end{equation}
We introduce a boundary current density $\mj^\mu$ as a Lagrange multiplier field that ensures continuity of $A$ over the boundary:\footnote{Note that $\mj^\alpha$ is a contravariant tensor \emph{density} of weight $+1$, as with our definition it already includes the metric pullback $\sqrt{-h}$ factor on $\partial \mathcal{M}$. We use this definition to match more naturally with the canonical formalism in curved space, where conjugate momenta are tensor densities.}
\begin{equation}
\prod_{x \in \partial \mathcal{M}} \delta\left(A_\mu-\bar{A}_\mu\right) = \int \left[\mathcal{D}\mathcal{J}^\mu\right] e^{i\int_{\partial \mathcal{M}}d^{d-1}x\, \mj ^\mu ( A_\mu-\bar{A}_\mu)}\label{mwdeltapij}
\end{equation}
The total action is then:
\begin{align}
\nonumber S= S_A + S_{\bar{A}} =&-\frac{1}{4}\int _{\mathcal{M}}d^d x\sqrt{-g}\,F^{\mu\nu}F_{\mu\nu}\\&+\int_{\partial \mathcal{M}}d^{d-1}x\, \mj ^\mu ( A_\mu-\bar{A}_\mu)-\frac{1}{4}\int _{\bar{\mathcal{M}}}d^d x\sqrt{-g}\,\bar{F}^{\mu\nu}\bar{F}_{\mu\nu},
\end{align}
and the path integral is now over $A_\mu,\bar{A}_\mu$ and $\mj^\mu$.

We obtain Maxwell theory in the submanifold $\mathcal{M}$ by getting rid of all the $\bar{A}$ contributions. The current $\mj$ is then to be interpreted as an external boundary source for Maxwell theory that is summed over. 

\begin{figure}[h]
\centering
\includegraphics[width=0.25\textwidth]{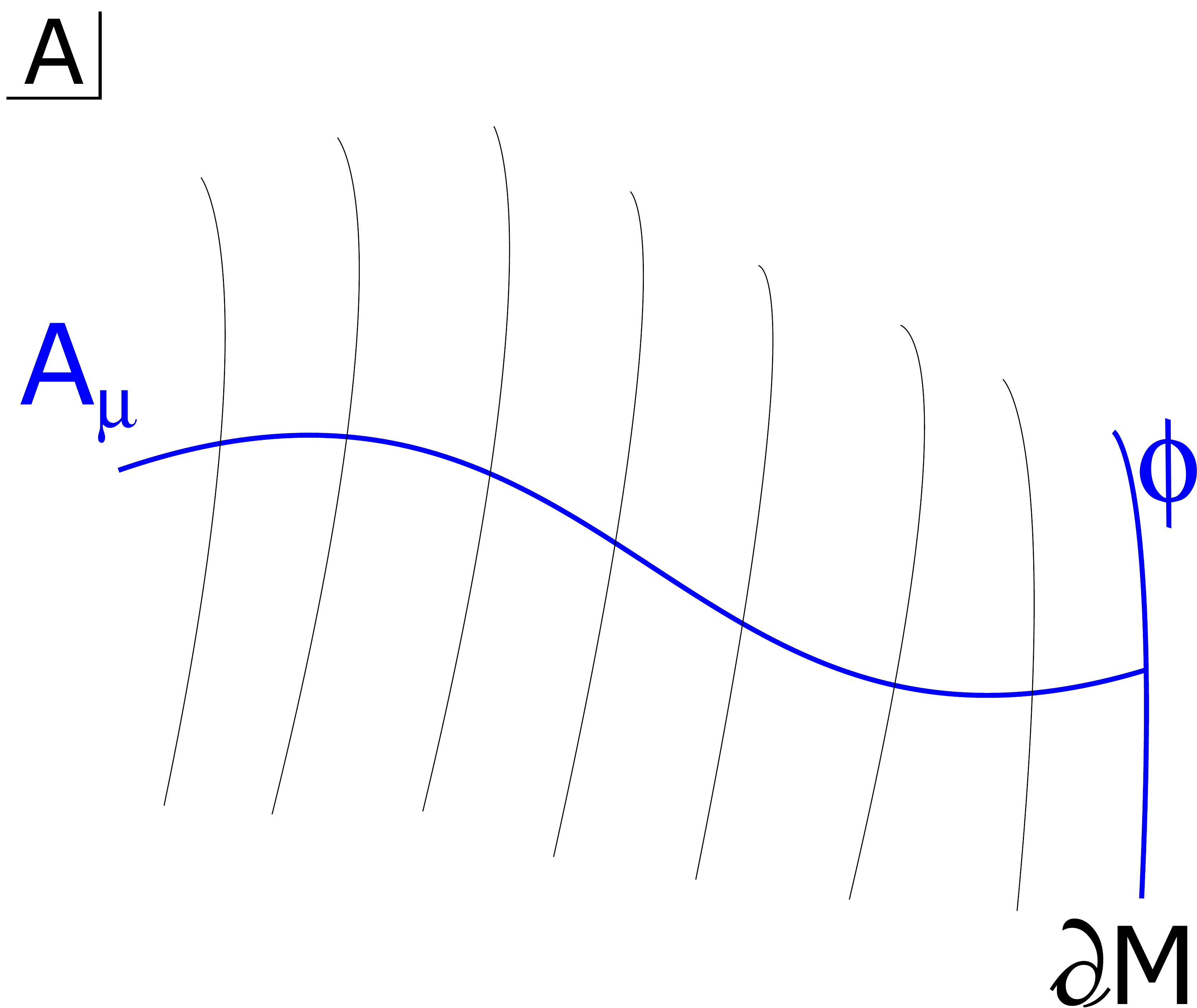}
\caption{Cartoon of the $A$ functional space. The blue lines are the physical fields. The bulk field is gauge-fixed to a single copy per gauge orbit. At the boundary the full would-be gauge orbit is rendered physical, parametrized by the gauge parameter $\phi$.}
\label{gaugeorbit}
\end{figure}

From the perspective of either side, the presence of the boundary breaks (large) gauge invariance, as the surface current $\mathcal{J}^\mu$ is introduced merely as a Lagrange multiplier, and is a priori not conserved. Large gauge transformations hence become dynamical (physical) fields on the boundary. To implement this restoration of degrees of freedom on the boundary, we write the gauge field as $A_\mu+\partial_\mu \phi$, where $A_\mu$ is considered fully gauge-fixed in the bulk $\mathcal{M}$ and where the only physical part of the field $\phi$ is living on the boundary surface $\partial \mathcal{M}$ (Figure \ref{gaugeorbit}).\footnote{We will show further on that $\partial_n \phi$ does not figure in the action, and hence only the variables $\left.\phi\right|_{\partial \mathcal{M}}$ and $\left.\partial_\alpha \phi\right|_{\partial \mathcal{M}}$ appear. }
The glued theory obtained by reintroducing $\bar{A}$ and integrating over $\mj$ obviously restores gauge-invariance on $\partial\mathcal{M}$. An alternative perspective on this construction can be found in appendix \ref{sect:fiber}, where we comment further on re-gluing the left- and right-sector.

Making this decomposition and acknowledging that on the boundary surface $\partial\mathcal{M}$ the would-be (large) gauge transformation $\phi$ becomes a physical degree of freedom, we are led to the action:
\begin{equation}
S_A=-\frac{1}{4}\int_\mathcal{M}d^d x\sqrt{-g}\, F^{\mu\nu}F_{\mu\nu}+\int_{\partial\mathcal{M}}\mj^\mu (A_\mu+\partial_\mu \phi)+S_\text{gf},\label{mwactiontotal}
\end{equation}
to be inserted into the path integral:
\begin{equation}
\label{Gpartial}
Z= \int \frac{\left[\mathcal{D} A_\mu\right]}{\text{vol }G} \, \frac{\left[\mathcal{D} \phi\right]}{\text{vol }G_\partial} \left[\dpi\mj\right] e^{iS_A}
\end{equation}
The path integral measure over the gauge field is divided by the volume of the gauge group $G$. From the perspective of either side of the boundary, the path integral also contains a division by the volume of the would-be gauge group \emph{at} the boundary to ensure a proper gluing is made when recombining both sides. The volume of the gauge group at the boundary is:
\begin{equation}
\text{vol }G_\partial = \int \left[\mathcal{D} \phi\right] = \delta(0),\label{volgd}
\end{equation}
which is just the functional generalization of $\lim_{k\to 0} \delta(k) = \lim_{k\to 0} \frac{1}{2\pi}\int_{-\infty}^{+\infty}\, dx \,e^{ikx}$. We will encounter it in this way later on.

For completeness we introduced the possibility of a gauge-fixing term $S_\text{gf}$ in the action arising from the Faddeev-Popov procedure, which will be irrelevant for our purposes (it vanishes on-shell).

The large gauge field $\phi$ that was introduced in \eqref{mwactiontotal} is only defined modulo constant shifts: $\phi \sim \phi + \text{constant}$, as its only goal in life is to parametrize the full would-be gauge orbit. For the $U(1)$ gauge group at hand these constant shifts are the global group $G = U(1)$ itself. This will be generalized to arbitrary compact groups in section \ref{sect:YM} below.
\\~\\
To proceed, we path integrate the system with action \eqref{mwactiontotal} over the bulk $A$-field. As the theory is quadratic, one evaluates the path integral by multiplying the determinant of quadratic fluctuations by the exponential of the on-shell classical action. The first is identified with the bulk photon degrees of freedom and the latter with the contribution from the boundary:
\begin{equation}
Z[\mathcal{J},\phi] = \int \left[\mathcal{D} A_\mu\right] e^{i S_A} = \left(\det \mathcal{O}\right)^{-\frac{1}{2}}\times e^{i \left.S_{A}\right|_\text{on-shell}} = Z^{\text{bulk}}\times Z^{\text{bdy}}[\mathcal{J},\phi].\label{zjphi}
\end{equation}
The determinant of the operator $\mathcal{O}$ might be difficult to evaluate explicitly in a curved spacetime, but it is associated to the $d-2$ bulk photon degrees of freedom, and not of interest for our purposes. It represents the bulk photon subject to perfectly magnetic conducting (PMC) boundary conditions:
\begin{equation}
Z^\text{bulk} =  \int_{\text{PMC}} \left[\mathcal{D} A_\mu\right]e^{iS_A} = \left(\det \mathcal{O}\right)^{-\frac{1}{2}}, \qquad \left.n_\mu F^{\mu\nu}\right|_{\partial \mathcal{M}} = 0, \quad \left.n_\mu A^\mu\right|_{\partial \mathcal{M}} = 0\label{mwzbulk}
\end{equation}
To isolate the boundary theory, we now simply strip off the fluctuation determinant. The result is an action where $A\left[\mj\right]$ is evaluated on-shell and where it represents the particular solution to the boundary conditions
\begin{equation}
\left.\left(\sqrt{-g} n_\mu F^{\mu\nu}\right)\right|_\text{bdy}= \mj^\nu, \label{mwbc}
\end{equation}
and the equations of motion
\begin{equation}
\nabla_\mu F^{\mu\nu}=0,\label{mweom1}
\end{equation}
that is orthogonal to the bulk photon. Notice that \eqref{mwbc} is inconsistent if $\mj_n \neq 0$, projecting the path integral on the $\mj_n=0$ sector only.

There is now a linear isomorphism between $\mj$ and $A$. Indeed, the boundary condition \eqref{mwbc} determines $F\rvert_\text{bdy}$ in terms of $\mj$. Since $A$ is completely gauge-fixed, this implies a linear isomorphism between $A\rvert_\text{bdy}$ and $\mj$. Evolving $A\rvert_{\text{bdy}}$ into the bulk using the equations of motion \eqref{mweom1} uniquely determines $A$ in the bulk: we denote the solution as $A[\mj]$. This procedure is unambiguous since $A$ decouples from the bulk photon because Maxwell theory is linear.
\\~\\
It is now straightforward to read off an action for the boundary variables: one evaluates the total action \eqref{mwactiontotal} on-shell using the procedure described above:
\begin{equation}
S^\text{bdy}\left[\mj,\phi \right]=\left.S_A\left[\mj,\phi,A\right]\right|_{\text{on-shell}}.\label{sbdysonshell}
\end{equation}
The result is an action quadratic in the boundary currents supplemented with a term coupling $\mj$ to the large would-be gauge field $\phi$. On-shell evaluation of the first part of \eqref{mwactiontotal} using integration by parts and the boundary condition \eqref{mwbc} results in 
\begin{equation}
-\frac{1}{2}\int d^d x \sqrt{-g}\,F^{\mu\nu}\partial_\mu A_\nu= -\frac{1}{2}\int d^{d}x\,\partial_\mu\left(\sqrt{-g}F^{\mu\nu}A_\nu\right)=-\frac{1}{2}\int_{\partial\mathcal{M}}d^{d-1}x\,\mj^\alpha A[\mj]_\alpha, \label{mwpart1eval}
\end{equation}
where we already used the fact that on-shell $\mj^n=0$. Obviously this only depends on the completely gauge-fixed part of $A$, since $F$ is manifestly gauge-invariant. The second part of \eqref{mwactiontotal} is:
\begin{equation}
\int_{\partial\mathcal{M}}d^{d-1}x\,\mj^\alpha \left(A[\mj]_\alpha+\partial_\alpha \phi\right).\label{japhi}
\end{equation}
In both these contributions, $A(\mj)$ is to be understood as an explicit linear function of $\mj$ obtained by on-shell evaluation. Summing these one obtains the Lorentzian boundary action:\footnote{A technicality is due here. The solution of the classical equations of motion enforces current conservation, so there is secretly a delta-functional present in a path integral over \eqref{mwsbdyjphi} enforcing this. However, this delta-functional is equally written as $\delta(0)$ without changing the value of the path integral as path-integrating over $\phi$ imposes the same constraint. In the notation of \eqref{zjphi}:
$$
Z^{\text{bdy}}[\mj,\phi]=\delta(0)e^{-S^\text{bdy}[\mj,\phi]}.
$$
This $\delta(0)$ then is canceled in the boundary path integral with the volume of the boundary gauge group $\text{vol }G_\partial$ as mentioned above in \eqref{Gpartial}. The result is \eqref{mwzbdy}.}
\begin{equation}
\boxed{S^\text{bdy}\left[\mj,\phi\right]=\int d^{d-1}x\, \left(\frac{1}{2}\mj^\alpha A[\mj]_\alpha + \mj^\alpha \partial_\alpha\phi\right)}.\label{mwsbdyjphi}
\end{equation}
Writing down this action makes manifest that the large gauge field $\phi$ has been promoted to a dynamical (physical) field variable living on the boundary, that is path-integrated over.\footnote{Small gauge transformations are irrelevant; they are captured by the bulk theory and are still being modded out of the physical Hilbert space by the usual Faddeev-Popov procedure.}

This action is in general higher-derivative. We initiate a canonical analysis in appendix \ref{app:constraint}.\footnote{
In detail, one solves the Maxwell problem \eqref{mweom1}, \eqref{mwbc} with a delta-source on the boundary to obtain the relevant Green function $G_{\alpha\beta}(y,x)$. One can then write the action explicitly:
\begin{equation}
\label{explact}
S^\text{bdy}\left[\mj,\phi\right]=\int d^{d-1}x\, \left(\frac{1}{2}\int d^{d-1}y \mj^\alpha(x) G_{\alpha\beta}(y,x) \mj^\beta(y) + \mj^\alpha \partial_\alpha\phi\right),
\end{equation}
which is non-local, but Gaussian.
} 
For all examples we will discuss further in the main text though, we will not need the generic case. From \eqref{mwsbdyjphi} one reads off the canonically conjugate field of $\phi$ as $\pi_\phi=\mj^t=\mathcal{Q}$, reminiscent of the analysis at the level of the presymplectic potential of \cite{Donnelly:2016auv}. In other words $\mathcal{Q}$ and $\phi$ have a nonvanishing Poisson bracket (or commutator in the quantum theory): $\comm{\phi}{\mathcal{Q}}=i$, or
\begin{equation}
\comm{\mathcal{Q}}{g}=-i g,\label{mwcommQg}
\end{equation}
where we exponentiated the element $\phi$ of the $U(1)$ algebra to obtain an element $g=e^\phi$ of the gauge group $U(1)$. The fact that we obtained the correct canonical structure of the theory directly at the level of the action was one of our main motivations.

The thermal partition function of the edge degrees of freedom is obtained by path-integrating the Euclidean version of \eqref{mwsbdyjphi} over the variables $\mj$ and $\phi$:
\begin{equation}
\boxed{Z^\text{bdy}=\int \left[\mathcal{D} \mj^\alpha\right]\left[\dpi \phi\right] e^{-S^\text{bdy}\left[\mj,\phi\right]}},\label{mwzbdy}
\end{equation}
where now the thermal action has been used:
\begin{equation}
S^\text{bdy}\left[\mj,\phi\right]=-\int_0^\beta d\tau \int d^{d-2}x\left(\frac{1}{2}\mj^\alpha A[\mj]_\alpha + \mj^\alpha \partial_\alpha\phi\right),\label{mwsEuclidean}
\end{equation}
with $\tau$ the coordinate along the thermal circle of circumference $\beta$ and the fields constrained to be periodic in $\beta$. In light of \eqref{mwcommQg}, (part of) this partition function is just a phase space path integral. We will come back to this further on.

The integration space of $\phi$ is all fields $\phi$ satisfying $\phi(\tau) = \phi(\tau+\beta) + \frac{2\pi}{e}n$ modulo $\phi \sim \phi + \text{constant}$ and $n\in \mathbb{Z}$. This is the loop group modulo the global group: $LG/G$.\footnote{The quantity $e$ denotes the fundamental charge, and the group element $g=e^{i e \phi}$ requires only periodicity of $\phi$ mod $\frac{2\pi}{e}$. This causes charge quantization, the charges being proportional to $\dot{\phi}$. For practical computations, we will ignore this by effectively setting $e\to 0$, leading to only $\dot{\phi}(\tau) = \dot{\phi}(\tau+\beta)$. The results can be readily adjusted to incorporate charge quantization. In effect, we take the gauge group to be $\mathbb{R}$ instead of $U(1)$.} This observation will be generalized to arbitrary compact groups later on.
\\~\\
There are two conjugate perspectives on computing the resulting phase space path integral. Firstly, the path integral over $\phi$ can be performed explicitly and results in a delta-functional on current conservation:
\begin{equation}
\label{sumQ}
Z^\text{bdy}=\int \left[\mathcal{D} \mj^\alpha\right]\delta(\partial_\alpha \mj^\alpha)e^{-S^\text{bdy}\left[\mj\right]},
\end{equation}
with the quadratic action for $\mj$ just the first part of \eqref{mwsEuclidean}. This immediately demonstrates the edge partition function is a summation over all possible current distributions on the boundary surfaces, weighted by a suitable action. \\

Alternatively, since $A[\mj]$ is linear in $\mj$ we can perform the Gaussian path integral over $\mj$ in \eqref{mwsEuclidean} to obtain a quadratic action for the dynamical field $\phi$:\footnote{Formally, this can be done very explicitly using \eqref{explact}.}
\begin{equation}
\boxed{Z^\text{bdy}=\int \left[\mathcal{D} \phi\right]e^{-S^\text{bdy}\left[\phi\right]}}.\label{mwgaugeaction}
\end{equation}
This is probably the most interesting view on the boundary edge theory: we obtain the action that governs the dynamics of the large would-be gauge degrees of freedom $\phi$.
\\~\\
How to re-obtain the full result by gluing back both theories with boundary is an interesting question but it is logically distinct from our main story. We provide details on this in Appendix \ref{sect:gluing}, with 2d examples that will be studied in the main text as well.
\\~\\
The remainder of this section consists of applications of these formulas to interesting examples. In sections \ref{s:warmup} and \ref{s:rindler}, we will discuss the evaluation of the edge partition function by explicitly evaluating \eqref{sumQ}. We will first apply it to an infinite plane in cartesian coordinates, and then take the specific example of Rindler space with the boundary at the horizon, a surface of infinite redshift. These two cases are interesting to compare, and we will confirm that this procedure produces the correct edge partition function. \\
In section \ref{s:2dmw} we specialize to 2d, as an example where the path integral over $\mj$ can be performed explicitly and one obtains a boundary theory with an explicit action for the pure gauge degrees of freedom. By comparing with the known literature, we show that the resulting boundary path integral produces the correct partition function.

\subsection{Application I: Maxwell Edge States in Flat Space}
\label{s:warmup}
As a warm-up and a first application, we consider an infinite plane in flat space and evaluate \eqref{sumQ} directly.
\begin{figure}[h]
\centering
\includegraphics[width=0.35\textwidth]{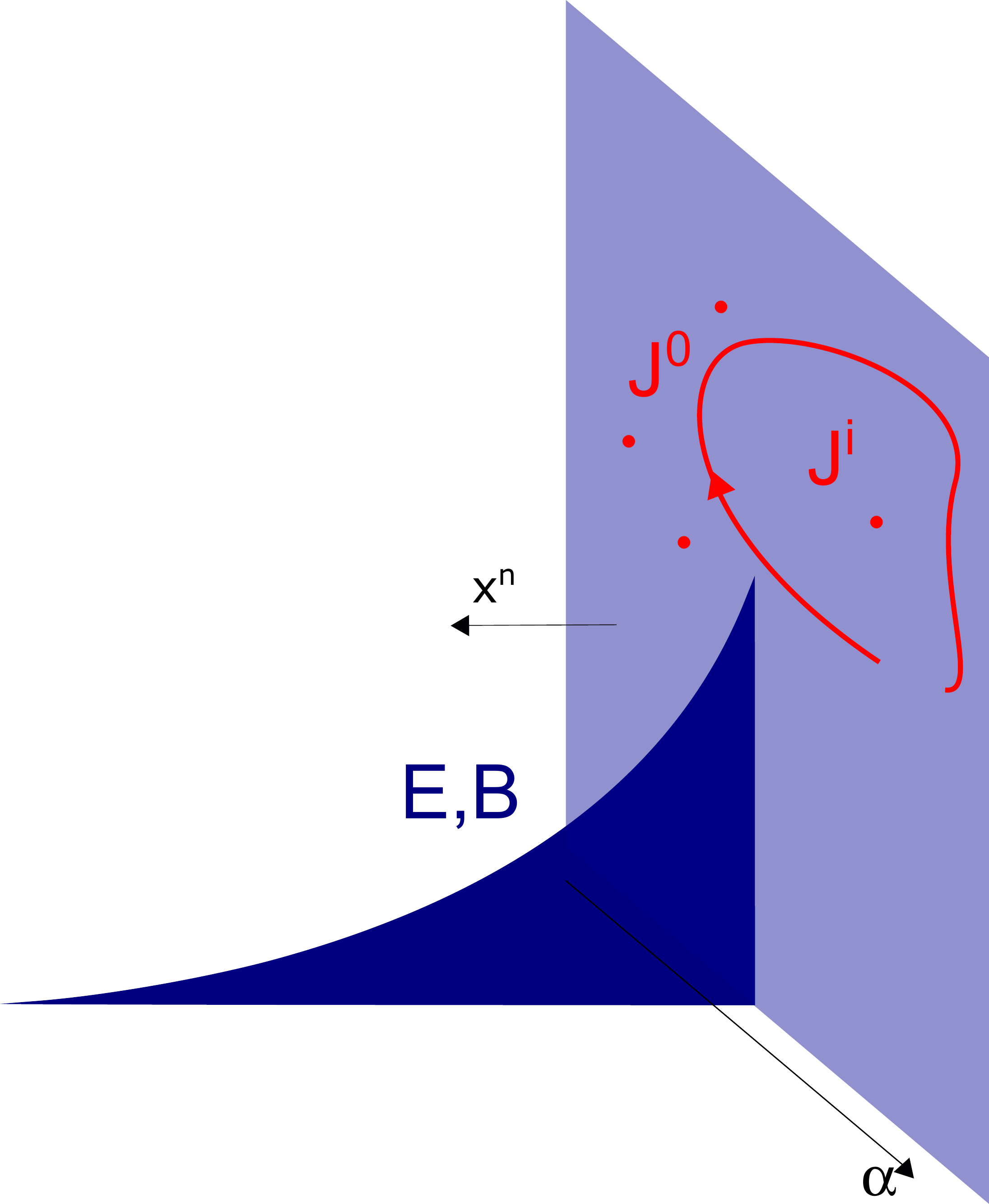}
\caption{Plane dividing space in two halves. Surface charges and currents source bulk electric and magnetic fields.}
\label{MWplane}
\end{figure}
We look for Maxwell solutions solving the bulk homogeneous equation $\Box A_\mu = 0$ and Lorenz gauge $\partial^\mu A_\mu = 0$, with boundary condition \eqref{mwbc}:
\begin{equation}
\left.\partial_n A_\alpha - \partial_\alpha A_n\right|_{x^n=0} = \mj_\alpha = \sum_k \mj_\alpha^k e^{i k \cdot x}.
\end{equation}
This classical Maxwell problem is solved by
\begin{equation}
A[\mj]_\alpha = \sum_{k} \frac{e^{i k \cdot x}}{ik_n} \mj^k_\alpha, \quad A_n = 0,
\end{equation}
with the sum over $k$ restricted by $\Box A_\mu=0$ to $k^2 = 0 = k_0^2 - \mathbf{k}^2-k_n^2$. As such, the sum is only over $k_0$ and $\mathbf{k}_\perp$ with $k_n^2 = k_0^2 - \mathbf{k}_\perp^2$. One readily checks Lorenz gauge is satisfied due to boundary current conservation $\partial_\alpha \mathcal{J}^\alpha = 0$.
\\~\\
This combined gauge $\nabla^\mu A_\mu=0$ and $A_n=0$ for the particular solutions $A[\mj]$ will be used for all computations in this work. It is a close relative to the standard radiation gauge, and it is particularly suited for our specific situation.
\\~\\
The on-shell Lagrangian is expanded as
\begin{equation}
\label{invopflat}
\frac{1}{2}\mj^\alpha A[\mj]_\alpha = \frac{1}{2} \mj^\alpha \frac{1}{\partial_n} \mj_\alpha = \frac{1}{2} \sum_{k,k'} \frac{\eta^{\alpha\beta}\mj^k_\alpha\mj^{k'}_\beta}{ik_n}e^{i(k-k')\cdot x}.
\end{equation}
Note that the sector $k_n=0$ does not contribute to the boundary partition function, because it carries an infinite energy. The boundary partition function is then
\begin{equation}
\label{Zflat}
Z^\text{bdy}=\int \left[\mathcal{D} \mj^\alpha\right]\delta(\partial_\alpha \mj^\alpha)e^{-\beta (2\pi)^{d-2}\sum_k \frac{(\mj^k_\alpha)^2}{2ik_n}} = \int \left[\mathcal{D} \mj^\alpha\right]\delta(\partial_\alpha \mj^\alpha)e^{-\beta (2\pi)^{d-2} \sum_{k} \frac{\Im(\mj^k_\alpha)^2}{\sqrt{k_0^2 - \mathbf{k}_\perp^2}}},
\end{equation}
where the last equality is found by rewriting the sum only over $k_n > 0$.\footnote{In the case that $k_0 < \mathbf{k}_\perp$, the integrand should be interpreted as
\begin{equation}
e^{-\beta (2\pi)^{d-2} \sum_{k} \frac{\Re(\mj^k_\alpha)^2}{\sqrt{\mathbf{k}_\perp^2-k_0^2}}},
\end{equation}
where in this case $k_n > 0$ which corresponds to an evanescent wave damped in the $x^n$-direction away from the boundary plane.} We path integrate over all charges $\mj_0$ and currents $\mj_i$ on the boundary surface, respecting boundary current conservation. \\
Note though that this is \emph{not} a manifest state counting interpretation of the Hilbert space, unlike \eqref{WDpf}, but just the thermal manifold evaluation of the partition function. This will be different in the next section \ref{s:rindler} when we discuss the Rindler case. We perform a preliminary canonical analysis of this system in appendix \ref{s:flkernel}.

\subsection{Application II. Maxwell Edge States in Rindler}
\label{s:rindler}
As a second explicit example, we consider Maxwell theory in Rindler space: $\mathcal{M}$ is the R-wedge of $\mathbb{R}^{1,d-1}$ and the boundary $\partial\mathcal{M}$ is the null Rindler horizon. We shall explicitly construct the boundary path integral \eqref{mwzbdy}. A first consistency check on the action \eqref{mwsbdyjphi} follows from a precise quantitative agreement with the edge partition function first obtained by Donnelly and Wall in \cite{Donnelly:2014fua,Donnelly:2015hxa}. As a bonus, the boundary action is shown to provide a convincing argument for the absence of static tangential magnetic fields (i.e. spatial boundary currents $\mj_i$) on the horizon as edge states, in contrast to \eqref{Zflat}.
\\~\\
Since the horizon is an infinite redshift surface, all fields living on it are necessarily static. One could for example infer this by demanding uniqueness of an arbitrary field at the horizon from the Euclidean perspective. This constrains the allowed currents in \eqref{mwsEuclidean} to be static.

\begin{figure}[h]
\centering
\includegraphics[width=0.3\textwidth]{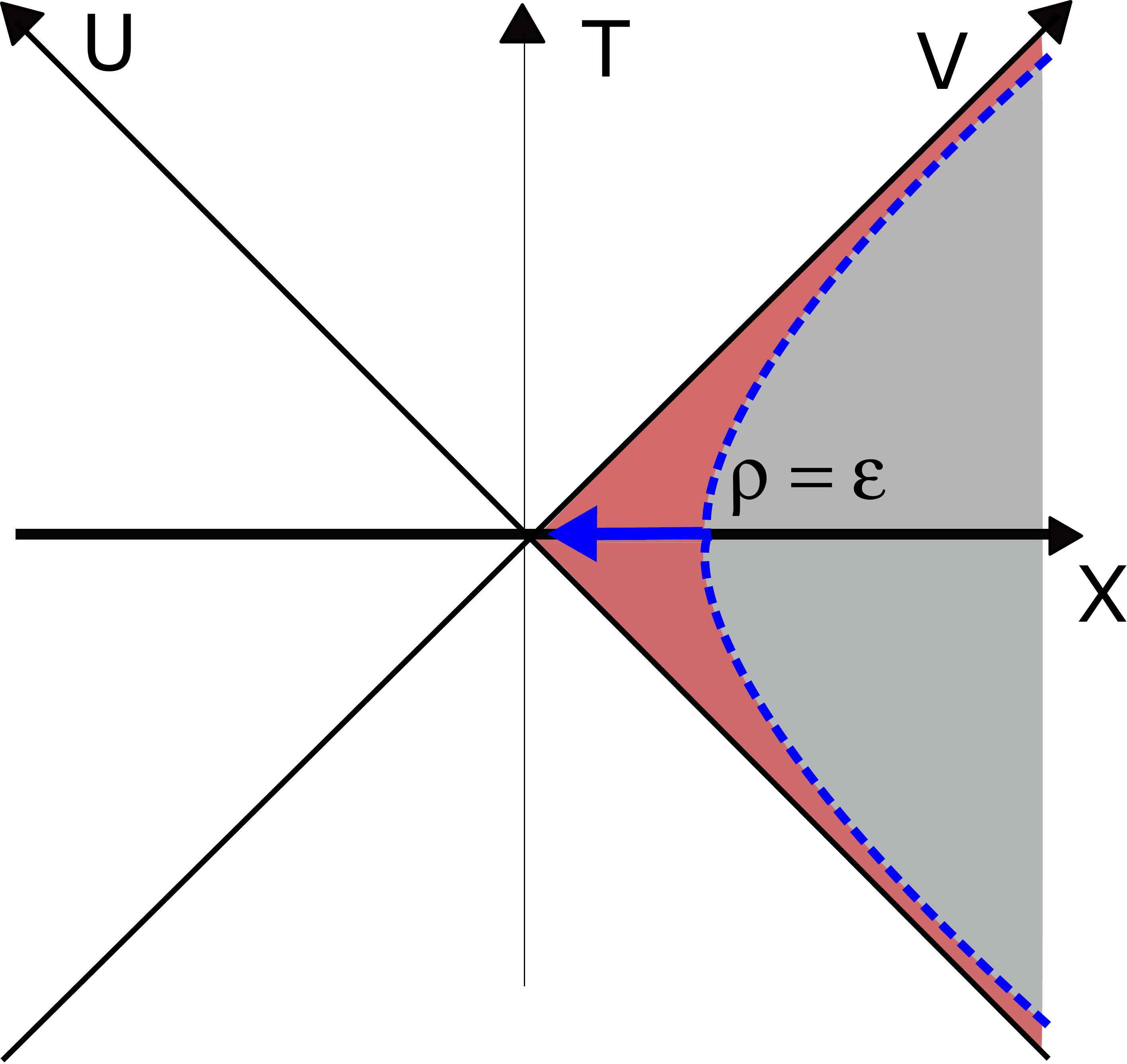}
\caption{Rindler space embedded in Minkowski space. The dashed blue line represents the regularized horizon (as a brick wall) that is taken in the limit to the actual null horizon.}
\label{Rindlerbrick2}
\end{figure}

To obtain an explicit boundary action \eqref{mwsbdyjphi} one searches for the bulk fields $F[\mj]$ and $A[\mj]$ which are isomorphic to the boundary current $\mj$. This is merely the equivalent of solving a Laplace equation with boundary conditions. At this point it is necessary to specify the the gauge: the isomorphism between $\mj$ and $A$ only holds when $A$ is completely gauge-fixed. We choose to implement the covariant Lorenz gauge:
\begin{equation}
\nabla^\mu A_\mu =0\label{mwgauge}.
\end{equation}
The Rindler coordinate system in $d$ dimensions is the metric:
\begin{equation}
ds^2=-\rho^2 dt^2+d\rho^2+{d\mathbf{x}}^2 = e^{2r}(-dt^2+dr^2)+d\mathbf{x}^2,\label{rindmetric}
\end{equation}
where $\rho = e^{r}$, and $\mathbf{x}=\{x^i\}$ denotes the coordinates parallel to the horizon i.e. the $d-2$ spectator dimensions. The boundary is chosen at $\rho=\epsilon \to 0^+$ as in Figure \ref{Rindlerbrick2}.

The details of the calculation are exiled to appendix \ref{app:maxrind}, not to distract from the main story. One obtains:
\begin{equation}
\mj^tA[\mj]_t\rvert_\text{bdy}=\mq\frac{1}{s\Delta}\mq,\label{jat}
\end{equation}
where $s\equiv -\ln \epsilon$ is a large positive regulator equal to (minus) the tortoise coordinate of the horizon that ultimately is to be taken to infinity, and $\Delta$ is the Laplacian on the $d-2$ spectator dimensions with negative eigenvalues $-k^2$. In addition one finds:
\begin{equation}
\mj^iA[\mj]_i \rvert_\text{bdy}=s\mj^i\mj^i,\label{jai}
\end{equation}
with summation implied. Inserting these into the action \eqref{mwsbdyjphi} one obtains:
\begin{equation}
\label{mxrii}
S\left[\mj,\phi\right]=\int d t \int d^{d-2}x \left(J^\alpha \partial_\alpha \phi -\mq\frac{1}{2s(-\Delta)}\mq+\frac{s}{2}\mj^i\mj^i\right).
\end{equation}
In the thermal partition function, the last term will contribute as $e^{s \frac{\mj^i\mj^i}{2}}$. Taking $s\to \infty$, the path integral will hence localize at the saddlepoint $\mj^i=0$. Notice that this $\mj^i=0$ localization proves unambiguously that there are no electromagnetic edge states associated with magnetic field configurations on the boundary, and the reason is the infinite horizon redshift. We are left with a phase space path integral over $\phi$ and $\pi_\phi=\mq$:
\begin{equation}
\boxed{Z=\int \left[\dpi \phi\right]\left[\dpi \pi_\phi\right]\exp{ -\int d\tau \int d^{d-2}x \left( \pi_\phi\frac{1}{2s (-\Delta)}\pi_\phi-i\pi_\phi \partial_\tau \phi\right)}}.\label{mwphasespace}
\end{equation}
The path integral over $\phi$ results in a factor $\delta(\dot{\pi_\phi})$ and one obtains:
\begin{align}
\label{thpiri}
Z=\int \left[\mathcal{D}\mq(\mathbf{x}) \right] \exp{ -\beta \int d^{d-2}x\, \mq\frac{1}{2s(-\Delta)}\mq}.
\end{align}
Due to the $\delta(\dot{\mq})$, this becomes a Gaussian functional integral over time-independent distributions $\mq(\mathbf{x})$ on the boundary. This path integral is precisely the edge partition function as first obtained by Donnelly and Wall in \cite{Donnelly:2014fua,Donnelly:2015hxa}, and studied in a canonical quantization context in \cite{Blommaert:2018rsf}. 
\\~\\
Further enforcing the limit $s\to\infty$ takes the action to zero, and we are left with:\footnote{An exception occurs when $\Delta=0$, i.e. when the charge distribution contains a spatial Fourier zero-mode, this is a spatial offset $\mq(\mathbf{x})=\mq + \delta \mq$. Such configurations have infinite action and are not counted.}
\begin{equation}
Z=\int \left[\mathcal{D}\mq(\mathbf{x}) \right].\label{mwcountingcharges}
\end{equation}
The boundary partition function is a simple counting of all electric charge configurations on the boundary with weight one; these are the electromagnetic microstates of the black hole, as pointed out in \cite{Blommaert:2018rsf}.

Of similar interest is the dual picture, where we obtain a boundary action for the large would-be gauge field $\phi$. Performing the path integral over $\pi_\phi$ in \eqref{mwphasespace} results in a quadratic boundary action for $\phi$:
\begin{equation}
\label{mwphi}
S\left[\phi\right]=\frac{s}{2}\int d^{d-1}x\, \partial_\tau \phi (-\Delta) \partial_\tau \phi.
\end{equation}
Enforcing $s\to\infty$ results in a localization on configurations for which $\dot{\phi}=0$. We are left with:\footnote{As stated earlier, the large gauge transformation $\phi$ is identified $\phi \sim \phi + \text{constant}$, so the counting that appears here is only over fields $\phi$ that have no overall spatial offset, matching the counting stated above in \eqref{mwcountingcharges}.}
\begin{equation}
Z=\int \left[\mathcal{D}\phi(\mathbf{x}) \right] \label{mwcountinggauge}.
\end{equation}
The boundary partition function is a simple counting of all time-independent large gauge transformations with weight one. This is the dual picture of \eqref{mwcountingcharges}.
\\~\\
The states of interest have zero energy as the above expressions demonstrate. This implies, in the canonical ensemble, that the entropy $S$ and partition function $Z$ are related as $S=\ln Z$, which is interpreted microcanonically as $Z = \Omega$, the number of states. \\
Both of the above formulas \eqref{mwcountingcharges} and \eqref{mwcountinggauge} can be rewritten in an alternative suggestive way, by using our old friend:
\begin{equation}
\int \left[\mathcal{D}\mq(\mathbf{x}) \right] = \prod_{\mathbf{x}} \int dq_x = \prod_{\mathbf{x}} \delta(0),
\end{equation}
to write
\begin{equation}
S = \ln Z = \int d\mathbf{x} \ln \delta(0) = A_H \ln \delta(0)
\end{equation}
which has manifest scaling with the area of the horizon $A_H$. The divergence $\ln \delta(0)$ reflects the UV-incompleteness of QFT near horizons. 
\\~\\
The resulting entropy $S$ can be viewed as residual entropy from the perspective of the Rindler thermal photon gas: the total energy of the photon gas is just the bulk piece (the thermal atmosphere of the black hole), but the total entropy contains both a bulk and an edge piece.
\\~\\
The fact that the on-shell evaluation only considers static configurations implies that \eqref{thpiri} coincides manifestly with a Hilbert space state-counting interpretation as mentioned in the Introduction. We elaborate on this, and the canonical structure of such a system in appendix \ref{s:tikernel}.
\\~\\
This result actually has a much wider applicability than expected. As usual, Rindler space is the near-horizon approximation to any black hole. So if the dynamics is confined to a region close to the horizon, Rindler space is a good approximation. For the edge theory however, the modes are stuck at the horizon: their wavefunction has no spread outside the horizon. This was illustrated in our earlier work \cite{Blommaert:2018rsf} (see e.g. Figure 10 therein), and implies that the Rindler approximation is exact to describe the edge sector for any non-extremal black hole.

\subsection{Application III. 2d Maxwell}
\label{s:2dmw}
As our next application and as a warm-up for what follows in section \ref{s:2dym}, we consider the special case of 2d Maxwell. The partition function of 2d Maxwell theory on a disk is well-known. Denoting by $A$ the area of the disk one obtains the integral
\begin{equation}
\label{ZMW}
Z=\int dEe^{-A\frac{E^2}{2}},
\end{equation}
where $E$ is the electric field normalized as $E^2=-\frac{1}{2}F^{\mu\nu}F_{\mu\nu}$. We can write this equivalently as a path integral over $E$ with a delta-constraint on static configurations:
\begin{equation}
\int dEe^{-A\frac{E^2}{2}}=\int\left[\dpi E\right]\delta(\dot{E})\exp{ -\int_0^\beta d\tau \frac{a}{2}E^2},
\end{equation}
where we introduced $a$ as $\beta a=A$. Introducing a Lagrange multiplier $\phi$ to replace the delta-functional, and replacing $E=\mq$, one obtains:
\begin{equation}
Z=\int \left[\dpi \mq \right]\left[\dpi \phi \right]\exp{ -\int_0^\beta d\tau \left(\frac{a}{2}\mq^2-i\mq \partial_\tau \phi\right)}.
\end{equation}
This is just the phase space path integral of a free particle, with $\pi_\phi=\mq$:
\begin{equation}
Z=\int \left[\dpi \pi_\phi \right]\left[\dpi \phi \right]\exp{-\int_0^\beta d\tau \left(\frac{a}{2}\pi_\phi^2-i\pi_\phi \partial_\tau \phi\right)}.
\end{equation}
This is made explicit by integrating out $\pi_\phi$, which results in the partition function of a particle on $U(1)$ with coupling $a$:
\begin{equation}
Z=\int d \mq e^{-A C_\mq}=\int \left[\dpi \phi\right]\exp {-\frac{1}{2a}\int d \tau \partial_\tau\phi \partial_\tau\phi},
\end{equation}
where $C_\mq=\mq^2/2$ is just the Casimir of $U(1)$, and the charge $\mq$ is the continuous parameter labeling the representations of $U(1)$. This label may become discrete due to charge quantization when additional matter fields are present, but we will not bother with this here. \\
Note that a particle on $U(1)$ is just a 1d massless scalar.
\\~\\
It is no coincidence that the partition function of the particle on $U(1)$ (with appropriate coupling) is precisely the partition function of 2d Maxwell on a disk. Indeed, using the logic of section \ref{s:bdy} we can directly reduce Maxwell on a disk to a particle on $U(1)$ living on the boundary of the disk. 

Consider 2d Maxwell on a Euclidean disk in polar coordinates $(\tau,\rho)$ with $\int d\rho \rho=a$. The bulk partition function of 2d Maxwell is just unity: the PMC boundary conditions restrict the theory to $F=0$, i.e. pure (small) gauge solutions, which are modded out. The boundary partition function is obtained as described above. With the boundary current just a charge $\mq$, the classical gauge field $A$ solution depends on the charge as $A[\mq]_t=-\frac{\rho^2}{2}\mq$. The Lorentzian boundary action \eqref{mwsbdyjphi} reduces to\footnote{As before, there is secretly a $\delta(\dot{\mq})$ present in the resulting path integral in order to solve the classical equations of motion. But $\phi$ also imposes this, so we replace it by $\delta(0)$ which cancels the volume of the boundary gauge group $G_\partial$ as before. The result is \eqref{2dmwz}.}
\begin{equation}
S\left[\mq, \phi\right]=\int dt  \left(-\frac{a}{2}\mq^2+\partial_t \phi \mq\right).\label{2dmwsbdy}
\end{equation}
The thermal boundary partition function hence becomes the phase space path integral:
\begin{equation}
Z=\int \left[\dpi \mq \right]\left[\dpi \phi \right]\exp{ -\int_0^\beta d\tau \left(\frac{a}{2}\mq^2-i\mq \partial_\tau \phi\right)},\label{2dmwz}
\end{equation}
which becomes a particle on $U(1)$ by integrating over $\mq$. This completes the proof.

\section{Yang-Mills}
\label{sect:YM}
It is not too hard to extend the discussion of section \ref{s:bdy} to Yang-Mills gauge theories with an arbitrary gauge group $G$. There is one important difference though: non-Abelian Yang-Mills theory is not free, as the field strength $F$ is nonlinear in $A$:
\begin{equation}
F^a_{\mu\nu}=\partial_\mu A_\nu^a-\partial_\nu A_\mu^a+\tensor{f}{^a_{bc}}A_\mu^b A_\nu^c.\label{fexpansion}
\end{equation}
This shows that there is no clean way to split bulk and boundary theories for non-Abelian Yang-Mills theory in general. Indeed, the theory is interacting and there will always be communication between different sectors. 

This being said, it's still interesting to investigate this boundary theory on its own. One reason for this is that the Yang-Mills boundary edge action will allow us to determine the canonical structure for the boundary, analogous to \eqref{mwcommQg}. A second reason is the special case of two-dimensional Yang-Mills discussed in section \ref{s:2dym}. This has no propagating bulk degrees of freedom and as such the theory is to a large extent determined by the edge theory. It is one of our goals to understand to which extent.

\subsection{Boundary Action for Yang-Mills}\label{s:ymba}
We denote the generators of the Lie algebra $\mathfrak{g}$ as $\tau^a$, satisfying the algebra $\comm{\tau^a}{\tau^b}=\tensor{f}{^{ab}_c}\tau^c$ and normalized as $\Tr(\tau^a\tau^b)=\delta^{ab}$. Following the same procedure as in section \ref{sect:Maxwell}, we introduce again a Lagrange multiplier field $\mj$ in the Lie algebra, and get the Lorentzian action:
\begin{equation}
S=-\frac{1}{4}\int_\mathcal{M}\Tr(F\wedge\star F)+\int_{\partial\mathcal{M}}d^{d-1}x\, \Tr (\mj^\mu A_\mu).\label{Gs}
\end{equation}
Under a generic large gauge transformation with group element $g=e^{\phi^a\tau_a}\in G$, the second part of the action transforms to 
\begin{equation}
\int_{\partial\mathcal{M}} d^{d-1}x\, \Tr(\mathcal{\mj}^\mu\left(g A_\mu g^{-1} - \partial_\mu g g^{-1}\right)), 
\label{Gbdyaction2}
\end{equation}
and the bulk action is gauge-invariant as usual. This formula is the analogue of \eqref{japhi}. Following the Maxwell discussion, to obtain the boundary action we will continue to use the form \eqref{Gbdyaction2} where it is now understood that $A$ is completely gauge-fixed, and $g$ represents the large gauge degrees of freedom: these have become physical (dynamical) degrees of freedom on the boundary.\footnote{As a consistency check, note that transferring between two different gauges for $A_\mu$ is contained within the field redefinition $g \to gh$ for a given specific field $h(x)$. The Jacobian of this transformation is trivial due to the left-invariance of the path-integral measure. For $U(1)$, the Jacobian of the transition $\phi \to \phi + \chi$ for a given $\chi(x)$ is even more quickly seen to be unity.} 

One needs to be careful when splitting the field into a gauge-fixed piece $A_\mu$ and a large gauge transformation $g$. Due to the fact that $A_\mu$ and $hA_\mu h^{-1}$ for a global (constant) gauge transformation $h$ are gauge-inequivalent, when writing the gauge field as $g A_\mu g^{-1} - \partial_\mu g g^{-1}$, the global transformations $h$ are already included as part of the gauge-fixed $A_\mu$ itself, and are not to be included in the $g$-integration. Such transformations would be obtained for $g$ when multiplying it on the right with $h$, so the path integral over the physical variable $g$ is over the right coset of the (product at every spatial point $\mathbf{x}$ of the) loop group: $\left(\prod_{\mathbf{x}}LG_{\mathbf{x}}\right)/G$ and excludes global transformations. This is the same conclusion as for the $U(1)$ case. 
Notice that the Lagrange multiplier current $\mj$ did not transform under the action of $g$.
\\~\\
As for Maxwell, we will emulate the path integration over the bulk gauge field $A$ by plugging in the classical equations of motion. Here however, as the action is quartic, there is no sensible splitting of bulk and edge degrees of freedom. The procedure we employ severs the interactions between bulk and boundary: the resulting theory will be interacting but only includes edge-edge interactions. Quantitatively, we only consider quadratic fluctuations around each non-perturbative saddle. As 2d Yang-Mills theory is by itself one-loop exact \cite{Witten:1992xu}, this covers the entire theory, as we show extensively in section \ref{s:2dym} further on. 
\\~\\
The Yang-Mills bulk equations of motion read:
\begin{equation}
(D_\mu F^{\mu\nu})^a=0,
\end{equation}
where $D$ is the covariant derivative including both the Christoffel connection $\Gamma$ and the gauge connection $A$. Variation of $A$ using the action \eqref{Gbdyaction} results in the boundary conditions:
\begin{equation}
g\left.\left(\sqrt{-g}n_\mu F^{\mu\nu}\right)\right|_\text{bdy}g^{-1}=\mj^\nu ,\label{ymbc}
\end{equation}
which implies $\mj^n=0$. This last constraint can be used to reduce the boundary action to:
\begin{equation}
\int d^{d-1}x\, \Tr(\mathcal{\mj}^\alpha\left(g A_\alpha g^{-1} - \partial_\alpha g g^{-1}\right)),\label{Gbdyaction}
\end{equation}
which is now internal to the boundary. The current $\mj$ can thus again be interpreted as a genuine boundary current sourcing the Yang-Mills equations. Notice that equation \eqref{ymbc} demonstrates the transformation behavior of $F$ under $g\to gh$: $\mj$ is invariant and hence $F$ transforms in the adjoint representation $F\to h^{-1}Fh$. 

The boundary action is the on-shell evaluation of the bulk action in the sense of \eqref{sbdysonshell}:
\begin{equation}
S^\text{bdy}\left[\mj,\phi \right]=\left.S_A\left[\mj,\phi,A\right]\right|_{\text{on-shell}}.
\end{equation}
Evaluation of the first term of the action \eqref{Gs} is analogous to the Maxwell evaluation \eqref{mwpart1eval}. One obtains the Lorentzian boundary action:
\begin{equation}
\boxed{S^\text{bdy}\left[\mj,g\right]=\int d^{d-1}x\, \Tr(\frac{1}{2}\mathcal{\mj}^\alpha g A[\mj]_\alpha g^{-1} - \mj^\alpha\partial_\alpha g g^{-1})}.\label{ymbdyaction}
\end{equation}
The boundary thermal partition function of Yang-Mills is thus:
\begin{equation}
Z^\text{bdy}=\int \left[\mathcal{D} \mj^\alpha\right]\left[\dpi g\right]e^{-S^\text{bdy}\left[\mj,g\right]},\label{Gzbdy}
\end{equation}
with the Euclidean boundary action
\begin{equation}
S^\text{bdy}\left[\mj,g\right]=-\int_{0}^{\beta}d\tau \int d^{d-2}x\, \Tr(\frac{1}{2}\mathcal{\mj}^\alpha g A[\mj]_\alpha g^{-1} - \mj^\alpha\partial_\alpha g g^{-1}),
\end{equation}
and where $gA[\mj]g^{-1}$ is now an explicit function of $\mj$ on account of the boundary condition \eqref{ymbc}. Notice that in general this action will not be quadratic in $\mj$ since $F$ is nonlinear in $A$. One way to obtain a boundary action quadratic in $\mj$ is to adopt radial gauge for the gauge field: $A_n=0$. On account of \eqref{fexpansion}, one now obtains a linear relation $g A[\mj]_\alpha g^{-1}\sim \mj_\alpha$ and hence a quadratic action \eqref{ymbdyaction}. Formally one could again path integrate out $\mj$ to obtain a boundary action $S^\text{bdy}\left[g\right]$ quadratic in the Maurer-Cartan one form $\omega=g^{-1}dg$. In section \ref{s:2dym} we will do this explicitly for 2d Yang Mills and obtain the action of a particle on $G$.
\\~\\
From \eqref{ymbdyaction}, one reads off the conjugate momentum of the group element $g$ as $\pi_g =  g^{-1}\mathcal{\mj}^t$. More particularly, element per element one defines
\begin{equation}
{\pi}_{ij}=\frac{\partial\mathcal{L}}{\partial \dot{g}_{ji}},\label{Gcancon}
\end{equation}
The gauge-invariant chromo-electric charges are defined using this conjugate momentum as $\mathcal{Q} = \mathcal{Q}^a \tau_a = g\pi_g = \mj^t$, or in components:
\begin{equation}
\mathcal{Q}^a=(g\pi_g)^a=\Tr(g\pi_g \tau^a)=(g\pi_g)_{ij}(\tau^a)_{ji}.\label{Gcharges}
\end{equation}
From the canonical algebra $\comm{g_{ij}}{\pi_{kl}}=i\delta_{il}\delta_{jk}$, inferred from \eqref{Gcancon}, we deduce the algebra of the charges \eqref{Gcharges}. For example:
\begin{equation}
\comm{\mathcal{Q}^a}{g_{kl}}=\comm{(g\pi_g)_{ij}(\tau^a)_{ji}}{g_{kl}}
=(\tau^a)_{ji}g_{im}\comm{\pi_{mj}}{g_{kl}}
=-i(\tau^a)_{ki}g_{il}=-i(\tau^a g)_{kl},
\end{equation}
or in short:
\begin{equation}
\boxed{\comm{\mathcal{Q}^a}{g}=-i\tau^a g}.\label{GcommQg}
\end{equation}
Likewise we deduce:\footnote{Some intermediate steps:
\begin{align}
\comm{\mathcal{Q}^a}{\mathcal{Q}^b}&=\comm{(g\pi_g)_{ij}(\tau^a)_{ji}}{(g\pi_g)_{kl}(\tau^b)_{lk}} \nonumber \\
&=(\tau^a)_{ji}(\tau^b)_{lk}\comm{g_{im}\pi_{mj}}{g_{ks}\pi_{sl}} \nonumber \\
&=g_{ks}\pi_{sj}\left((\tau^a)_{ji}(\tau^b)_{ik}-(\tau^a)_{ik}(\tau^b)_{ji}\right) \nonumber \\
&=\tensor{f}{^{ab}_c}(g\pi_g)_{kj}(\tau^c)_{jk} =\tensor{f}{^{ab}_c}\mathcal{Q}^c.
\end{align}
}
\begin{equation}
\boxed{\left[\mathcal{Q}^a, \mathcal{Q}^b\right] =\tensor{f}{^{ab}_c} \mathcal{Q}^c}.\label{GcommQQ}
\end{equation}
The relations \eqref{GcommQg} and \eqref{GcommQQ} confirm that the Lie algebra valued charges $\mathcal{Q}$ generate large would-be gauge transformations which are physical fields on the boundary.

This is precisely the canonical boundary algebra obtained by Donnelly and Freidel from an analysis of a necessary boundary contribution to the presymplectic potential \cite{Donnelly:2016auv}.\footnote{In \cite{Donnelly:2016auv}, the field $g$ is introduced as an external field that transforms under a large gauge transformation with element $h\in G$ as $g\to gh^{-1}$, and $A$ in \eqref{Gbdyaction} transforms under $h$ in the usual manner $A\to hAh^{-1}-dh h^{-1}$ . The coupling of the gauge field to the external current $\mj$ in \eqref{Gbdyaction} is then completely gauge-invariant. The action \eqref{Gbdyaction} would thus be the one that goes with their analysis and reproduces their canonical boundary structure. We are led instead to the interpretation that the field $g$ \emph{is} the gauge freedom on the boundary, which has become dynamical. Large gauge invariance is only restored upon gluing.}
The recovery of the correct boundary canonical structure \eqref{GcommQg} and \eqref{GcommQQ} directly from the boundary action \eqref{ymbdyaction} is an important consistency check on the validity of this construction.
\\~\\
As in section \ref{sect:Maxwell}, we provide examples to illustrate this construction.

\subsection{Application IV. 2d Yang-Mills}
\label{s:2dym}
An especially interesting application of the boundary partition function \eqref{Gzbdy} is to consider theories where there are no propagating bulk degrees of freedom. In these cases, one expects to be able to cleanly pinpoint the edge sector of the theory. As the simplest example, the edge construction of a purely topological theory is well-known, as the theory fully reduces to just this piece. A somewhat less trivial example is two-dimensional Yang-Mills theory: the theory is quasi-topological in the sense that generic correlators not only depend on the topology of the manifold, but also on the areas enclosed by Wilson lines on the manifold.
\\~\\
The logic of this work provides us with a way to associate a 1d boundary action to 2d Yang Mills with gauge group $G$. We will show this boundary theory to be a particle on the group $G$. We will match the partition function of a particle on a group on the boundary of a disk with that of 2d Yang Mills in the interior of the disk, thereby providing an important check on our methods.\footnote{It is not difficult to generalize this to arbitrary Riemann surfaces.} Correlators of these theories are of particular interest and are discussed separately in section \ref{s:cor}.
\\~\\
As in section \ref{s:2dmw}, we will first rewrite the known results of 2d Yang-Mills theory in a suggestive way. Afterwards we will provide the direct derivation of the edge action using our procedure from the previous section.

The Euclidean path integral of 2d Yang-Mills on a disk with area $A$ and monodromy $U \equiv P\, \exp{\oint_\mathcal{C}A}$ around the boundary $\mathcal{C}$ is:
\begin{equation}
Z(U)=\sum_R \dim R \, \chi_R (U)\, e^{-A C_{R}},\label{2dYMzU}
\end{equation}
where the sum ranges over all irreps $R$ of $G$, $\chi_R$ is the character in $R$, and $C_R$ is the Casimir in the irrep $R$. Choosing the trivial monodromy $U=1$ one obtains:
\begin{equation}
Z=\sum_R (\dim R)^2e^{-A C_{R}}.\label{2dYMz}
\end{equation}
Using techniques well-known within the coadjoint orbit literature \cite{Alekseev:1990mp,Alekseev:1988vx}, this can be rewritten as a double phase space path integral over $g \in LG/G$ and a Lie-algebra valued field $\mq$:
\begin{equation}
Z = \int \left[\dpi \mq\right]\left[\dpi g\right]\exp{-\int_0^\beta d\tau \Tr(\frac{a}{2}\mq^2+i\mq \partial_\tau g g^{-1})}.
\end{equation}
We provide details in appendix \ref{app:AS}.\footnote{Such an action was written down in the past in \cite{Alekseev:1993np} in a different context as a toy model for canonical quantization in systems with Poisson-Lie symmetry.}
Integrating out $\mq$ results in:
\begin{equation}
\boxed{Z = \int_{LG/G}\left[\dpi g\right]\exp{-\frac{1}{2a}\int d\tau \Tr(g^{-1}\partial_\tau g g^{-1}\partial_\tau g)}},\label{poag}
\end{equation}
which is the partition function of a particle on the group $G$ with coupling $a$.
\\~\\
As for the Maxwell case, this equality can alternatively be obtained directly using the boundary action \eqref{ymbdyaction}. The goal is to find the analogue of \eqref{jat} for 2d Yang-Mills. The Lorentzian boundary action \eqref{ymbdyaction} is:
\begin{equation}
S\left[\mq,g\right]=\int dt \Tr(\frac{1}{2}\mq g A(\mq)_t g^{-1}-\mq \partial_t g g^{-1}).\label{ymac1}
\end{equation}
The boundary conditions relate $F$ and $\mq$ by 
\begin{equation}
\left.\left(\sqrt{-g}n_\mu F^{\mu t}\right)\right|_\text{bdy}=g^{-1}\mq g,
\end{equation}
or 
\begin{equation}
g\left.\left(\sqrt{-g}n_\mu F^{\mu t}\right)\right|_\text{bdy}g^{-1}=\mq.\label{gfgq}
\end{equation}
In general, $F$ depends nonlinearly on $A$ through $F_{\mu\nu}^a=\partial_\mu A_\nu^a -\partial_\nu A_\mu^a+\tensor{f}{^a_{bc}}A^b_\mu A^c_\nu$, and as such a linear relation between $gA(\mq)g^{-1}$ and $\mq$ is not guaranteed by \eqref{gfgq}. Fortunately, we can obtain a particular solution that ensures precisely such a linear relationship which makes it possible to integrate out $\mq$.

The required solution is the precise equivalent of the Maxwell solution: $A_t^a=-\frac{\rho^2}{2}\mq^a$. Indeed, this field is a solution of the bulk equations of motion $(D_\mu F^{\mu\nu})^a=0$ and it satisfies the boundary conditions \eqref{gfgq}. It also satisfies the Lorenz gauge condition: $(D^\mu A_\mu)^a=\nabla^\mu A_\mu^a=0$. Some details are presented in Appendix \ref{app:class}. Inserting this solution in \eqref{ymac1} results in the Lorentzian boundary action:
\begin{equation}
S\left[\mq,g\right]=-\int dt \Tr(\frac{a}{2}\mq^2+\mq \partial_t g g^{-1}),\label{ymac2}
\end{equation}
which is the precise equivalent of \eqref{2dmwsbdy}. The thermal boundary partition function becomes just:
\begin{equation}
Z=\int \left[\dpi \mq\right]\left[\dpi g\right]\exp{-\int_0^\beta d\tau \Tr(\frac{a}{2}\mq^2+i\mq \partial_\tau g g^{-1})},
\end{equation}
which becomes a particle on $G$ by integrating over $\mq$.
\\~\\
We comment on gluing two such systems back together in Appendix \ref{app:YM}.
\\~\\
It is relatively straightforward to observe that the more general situation of 2d Yang-Mills on a disk with monodromy $U$ around the boundary curve \eqref{2dYMzU} is obtained by imposing twisted boundary conditions on the group element $g$: $g(\tau + \beta) = Ug(\tau)$. Indeed, the twisted partition function of the particle on a group:
\begin{equation}
\int dg K(Ug, g;\beta) = \text{Tr}(e^{-\beta H} U )= \sum_R \dim (R) \,\, \chi_R(U) e^{-\beta a C_{R}},
\end{equation}
becomes precisely \eqref{2dYMzU}.
\\~\\
The partition function of 2d Yang-Mills \eqref{poag} is that of a particle on $G$. The same is not true for an arbitrary correlation function; 2d Yang-mills is only quasi-topological: a closed Wilson line inserted deep in the 2d bulk affects the path integral of 2d Yang Mills, but not the edge theory. As we will highlight in section \ref{s:cor} though, the link between both theories goes much further than a mere equivalence on the level of partition functions: disk expectation values in 2d Yang Mills of a large subclass of boundary-anchored Wilson lines can all be calculated using particle-on-a-group correlators.

\subsection{Application V. Yang-Mills Edge States in Rindler}
\label{s:YMrindler}
The non-Abelian generalization of section \ref{s:rindler} is straightforward. On-shell evaluation $A[\mj]$ is identical to the Maxwell example because we can resort to $A_\rho=0$ gauge for the particular solutions in the $\omega=0$ sector. As explained below equation \eqref{ymbdyaction}, this results in a linear relation $A[\mj]\sim \mj$ making the edge action again quadratic in $\mj$, effectively reducing the Yang-Mills on-shell evaluation to $\text{dim }G$ copies of the Maxwell case. We obtain the analogue of \eqref{mxrii}:
\begin{equation}
\label{YMbefore}
S\left[\mj,g\right]=\int d t d^{d-2} \mathbf{x} \Tr \left(J^\alpha \partial_\alpha g g^{-1} -\mq\frac{1}{2s(-\Delta)}\mq+\frac{s}{2}\mj^i\mj^i\right).
\end{equation}
Taking $s\to\infty$, the path integral again localizes on $\mj^i=0$ and we are left with
\begin{equation}
Z=\int \left[\dpi g\right]\left[\dpi \mq\right]\exp{i\int d\tau d\mathbf{x}\Tr \left(\mq\partial_\tau g g^{-1}\right)}.\label{zsinftyym}
\end{equation}
Path integrating over $\mq$ results in a sum with unit weight over configurations $\partial_\tau g g^{-1}=0$ or time-independent $g(\mathbf{x})$:
\begin{equation}
Z=\int \left[\dpi g(\mathbf{x})\right].\label{ymcountinggauge}
\end{equation}
This is just summing static large gauge configurations on the boundary. 

An interesting perspective is obtained by integrating out $g$ and $\mq$ directly in \eqref{zsinftyym} using the techniques of appendix \ref{app:AS}. We obtain the analogue of \eqref{pipq} but now integrating over the angular variables results in space-dependent solutions $m_i^k(\mathbf{x})$, with zero Hamiltonian. The result is a sum over all states in all representations of $G$ at each spatial point $\mathbf{x}$ on the horizon:
\begin{equation}
Z=\prod_\mathbf{x}\sum_{R(\mathbf{x})} (\dim R(\mathbf{x}))^2,
\end{equation}
which is UV-divergent. Denoting the divergent dimension of the total state space for $G$ as $\Omega_G=\sum_R (\dim R)^2$, the residual entropy is
\begin{equation}
S=A_H\ln \Omega_G =A_H \ln \int d \tilde{g},
\end{equation}
where one has the formal equality of $\sum_R (\dim R)^2 = \int d\tilde{g}$, the volume of $\tilde{G}$, the universal cover of $G$. 
\\~\\
Both here as in section \ref{s:rindler}, our final result looks like the 2d theory result taken at every point of the Rindler horizon in the zero-temperature limit $\beta \to \infty$, cfr. \eqref{ZMW} for the Maxwell case, and \eqref{2dYMz} for the YM case, where $A=a\beta$. Both the fact that $\beta \to \infty$ and the transverse decoupling are due to infinite redshift. Firstly, every finite energy excitation at the horizon is redshifted to zero energy at the location of the Rindler (or Schwarzschild) observer. Secondly, separate points on the horizon cannot communicate with each other, and the edge computation effectively reduces to a 1+1d computation. We can see this explicitly when going from \eqref{YMbefore} to \eqref{zsinftyym}: the limit removes all $\mathbf{x}$-derivatives in the Lagrangian, making the theory ultralocal in the transverse $\mathbf{x}$-directions, and removing all correlation between different locations on the horizon. \\
This is an additional motivation for studying the quasi-topological two-dimensional cases.

\section{Edge Correlators in 2d Yang-Mills}
\label{s:cor}
The edge theory of 2d Yang-Mills on a disk describes a particle on a group on the boundary circle. In recent work \cite{Mertens:2018fds}, one of the authors calculated the correlators of the particle-on-a-group model by dimensionally reducing 2d Wess-Zumino-Witten (WZW) conformal field theory between vacuum branes.

In this section we relate these results to computations within 2d Yang-Mills and answer which aspects of the bulk YM theory are captured by correlators in just the boundary theory.  More in particular we will show which YM correlators can be calculated using particle-on-a-group correlators, and which can't. For the technical computations, we have in mind the gauge group $G=SU(2)$, but our expressions can equally be interpreted directly for any compact gauge group.

\subsection{Particle on a Group Correlation Functions}
Particle-on-a-group correlators can be obtained from WZW correlators by dimensionally reducing WZW between two vacuum branes, each characterized by a vacuum state that can be expanded in Ishibashi states. The details can be found in \cite{Mertens:2018fds}, we will only review the final results for the correlation functions that this procedure leads to.
\\~\\
The first thing to consider is which operator insertions to include in the particle-on-a-group path integral. As discussed in \cite{Mertens:2018fds} from the 2d WZW perspective, using the Peter-Weyl theorem, one can write the most generic local primary operator in WZW (in lightcone ($u,v$) coordinates) as a linear combination of the elementary operators:\footnote{Primary operators are constructed as functions of $g(u,v)$, but not its derivatives.} 
\begin{equation}
\mathcal{O}_{R,m\bar{m}}(u,v)=R(g(u,v))_{m\bar{m}},
\end{equation}
where $R$ is a certain irrep of $G$ and $m$, $\bar{m}$ are two labels in the irrep each ranging over $\dim R$ values. For example, for $G=SU(2)$ and $R=j$ this is just $m\in \{-j,-j+1,..,j\}$. Dimensional reduction from 2d WZW to 1d particle-on-a-group results in the bilocal operator:
\begin{equation}
\mathcal{O}_{R,m\bar{m}}(\tau_1,\tau_2)=R(g(\tau_2)g^{-1}(\tau_1))_{m\bar{m}}.\label{bilocal}
\end{equation}
Note that this bilocal operator is invariant under global $G$ transformations $g\to gh$. As the particle-on-a-group path integral \eqref{poag} has global $G$ transformations as a gauge symmetry, this means this bilocal operator is on its own already gauge-invariant, and an interesting observable to consider.

A 2d-1d holographic intuition into why these bilocals are so natural was not provided previously though. We will point out in what follows that they have a bulk interpretation in terms of Wilson lines in 2d Yang-Mills. As the theory is only quasi-topological, not all features of 2d YM are captured by these edge correlators. A more direct link between the particle-on-a-group model and its topological holographic dual: 2d BF theory, can also be given in terms of boundary-anchored Wilson lines in the BF bulk and is discussed elsewhere \cite{Blommaert:2018.2}.

Combining 2d CFT techniques with a doubled version of the Wigner-Eckart theorem, correlation functions of such bilocal operator insertions were determined \cite{Mertens:2018fds}. We next summarize the results of this computation.
\\~\\
The two-point function is the expectation value of a single bilocal operator $\mathcal{O}_{R,m\bar{m}}(\tau_1,\tau_2)$:
\begin{align}
\nonumber \left\langle \mathcal{O}_{R,m\bar{m}}(\tau_1,\tau_2)\right\rangle=&\sum_{R_1,R_2}\dim R_1\dim R_2 \,e^{-a(\tau_2-\tau_1)C_{R_2}}e^{-a(\beta-\tau_2+\tau_1)C_{R_1}}\\
&\cross \sum_{m_1,m_2}\tj{R_1}{R}{R_2}{m_1}{m}{m_2}\tj{R_1}{R}{R_2}{m_1}{\bar{m}}{m_2},\label{pg2pt}
\end{align}
where the $3j$-symbols of the group have been introduced. Using the identity 
\begin{equation}
\sum_{m_1,m_2}\tj{R_1}{R}{R_2}{m_1}{m}{m_2}\tj{R_1}{R}{R_2}{m_1}{\bar{m}}{m_2}=\frac{1}{\dim R}\delta_{m\bar{m}}N_{R_1RR_2},
\end{equation}
we can simplify \eqref{pg2pt} into:
\begin{align}
\left\langle \mathcal{O}_{R,m\bar{m}}(\tau_1,\tau_2)\right\rangle = \sum_{R_1,R_2}&\dim R_1\dim R_2 \, e^{-a(\tau_2-\tau_1)C_{R_2}}e^{-a(\beta-\tau_2+\tau_1)C_{R_1}}\frac{N_{R_1RR_2}}{\dim R}\delta_{m\bar{m}},
\end{align} 
which is diagonal in $m$ and $\bar{m}$: only the diagonal bilocals are non-zero. From the 2d WZW CFT perspective, the $m$ and $\bar{m}$ labels are associated to respectively holomorphic and antiholomorphic sectors; the diagonal operators are spinless in 2d CFT. 
\\~\\
The time-ordered four-point function is the expectation value of two bilocal operators, with $\tau_1<\tau_2<\tau_3<\tau_4$:
\begin{align}
\nonumber \Big\langle \mathcal{T}\mathcal{O}_{R_A,m_A\bar{m}_A}&(\tau_1,\tau_2) \mathcal{O}_{R_B,m_B\bar{m}_B}(\tau_3,\tau_4) \Big\rangle =\sum_{R_1,R_2,R_3}\prod_i \left(\dim R_i e^{-a L_i C_{R_i}}\right)\\&\cross\sum_{m_1,m_2,m_3,\tilde{m}_3}\tj{R_1}{R_A}{R_3}{m_1}{m_A}{m_3}\tj{R_1}{R_A}{R_3}{m_1}{\bar{m}_A}{\tilde{m}_3}\tj{R_2}{R_B}{R_3}{m_2}{m_B}{\tilde{m}_3}\tj{R_2}{R_B}{R_3}{m_2}{\bar{m}_B}{m_3},\label{pg4pt}
\end{align}
which is not necessarily diagonal. Here $L_i$ are the respective lengths of the boundary segments.
\\~\\
It is clear at this point that the exact answers for the correlators such as \eqref{pg4pt} are highly structured. As mentioned in \cite{Mertens:2018fds}, a diagrammatic decomposition can be used to write down the general amplitude.

\begin{itemize}
\item The starting point is the oriented thermal circle. A bilocal operator $\mathcal{O}_{R,m\bar{m}}(\tau_i,\tau_f)$ becomes an oriented line from $\tau_i$ to $\tau_f$ with label $R$. The starting point $\tau_i$ receives a label $m$, the endpoint receives the label $\bar{m}$. 
\item Each region in the resulting diagram is assigned an irrep $R_i$, and contributes a weight $\dim R_i$. Assign a $m_i$ label to each boundary segment. Eventually these labels $R_i$ and $m_i$ are to be summed over.
\item Each boundary segment carries a propagation factor $e^{- a L_i C_{R_i}}$, proportional to the length $L_i$ of the relevant segment $i$. Each intersection of an endpoint of an internal line with the boundary has associated with it 3 irreps and 3 labels and is weighed with the $3j$-symbol associated with these labels.
\begin{align}
\label{frules}
\begin{tikzpicture}[scale=0.8, baseline={([yshift=-0.1cm]current bounding box.center)}]
\draw[thick] (-0.2,0) arc (170:10:1.53);
\draw[fill,black] (-0.2,0.0375) circle (0.1);
\draw[fill,black] (2.8,0.0375) circle (0.1);
\draw (3.4, 0) node {\footnotesize $\tau_1$};
\draw (-0.7,0) node {\footnotesize $\tau_2$};
\draw (1.25, 1.6) node {\footnotesize $\textcolor{red}{m}$};
\draw (1.25, 0.5) node {\footnotesize $R$};
\draw (5, 0) node {$\raisebox{6mm}{$ = e^{- C_{R} (\tau_2-\tau_1)}$}$};
\end{tikzpicture} ~~~~~~~~~~\ \ \begin{tikzpicture}[scale=1, baseline={([yshift=-0.1cm]current bounding box.center)}]
\draw[thick] (-.2,.9) arc (25:-25:2.2);
\draw[fill,black] (0,0) circle (0.08);
 \draw[thick,blue](-1.5,0) -- (0,0);
\draw (.5,-1) node {\footnotesize $\textcolor{red}{m_2}$};
\draw (.5,1) node {\footnotesize$\textcolor{red}{m_1}$};
\draw (-0.3,.3) node {\footnotesize$\textcolor{red}{m}$};
\draw (-1,.3) node {\footnotesize$ \color{blue}R$};
\draw (-1,.8) node {\footnotesize$ R_1$};
\draw (-1,-.5) node {\footnotesize$ R_2$};
\draw (3,0.1) node {$\mbox{$\ =\  \, \tj{R_1}{R_2}{R}{m_1}{m_2}{m}.$}$}; \end{tikzpicture}\ 
\end{align}
\item Each crossing of two internal lines is associated with 6 irreps and is weighed with the appropriate $6j$-symbol, by the rule:
\begin{align}
\label{crossing}
\ \begin{tikzpicture}[scale=1, baseline={([yshift=0cm]current bounding box.center)}]
\draw[thick,blue] (-0.85,0.85) -- (0.85,-0.85);
\draw[thick,blue] (-0.85,-0.85) -- (0.85,0.85);
\draw[dotted,thick] (-0.85,-0.85) -- (-1.25,-1.25);
\draw[dotted,thick] (0.85,0.85) -- (1.25,1.25);
\draw[dotted,thick] (-0.85,0.85) -- (-1.25,1.25);
\draw[dotted,thick] (0.85,-0.85) -- (1.25,-1.25);
\draw (1.5,0) node {\scriptsize $R_4$};
\draw (-1.5,0) node {\scriptsize $R_2$};
\draw (-.75,.33) node {\scriptsize \color{blue}$R_A$};
\draw (.78,.33) node {\scriptsize \color{blue}$R_B$};
\draw (0,1.5) node {\scriptsize  $R_1$};
\draw (0,-1.5) node {\scriptsize $R_3$};
\end{tikzpicture}~~\raisebox{-3pt}{$\ \ \  = \ \ \sj{R_B}{R_1}{R_4}{R_A}{R_3}{R_2}$}~~~
\end{align}

\end{itemize}

\noindent As an example, the above two-point function \eqref{pg2pt} and four-point function \eqref{pg4pt} are diagrammatically:
\begin{align}
\begin{tikzpicture}[scale=1, baseline={([yshift=0cm]current bounding box.center)}]
\draw[thick] (0,0) circle (1.5);
\draw[thick,blue] (-1.5,0) -- (1.5,0);
\draw[fill,black] (-1.5,0) circle (0.1);
\draw[fill,black] (1.5,0) circle (0.1);
\draw (0,1.7) node {\small \color{red}$m_1$};
\draw (1.2,0.2) node {\small \color{red}$m$};
\draw (-1.2,0.2) node {\small \color{red}$\bar{m}$};
\draw (0,-1.7) node {\small \color{red}$m_2$};
\draw (-2,0) node {\small $\tau_2$};
\draw (2,0) node {\small $\tau_1$};
\draw (0,.25) node {\small \color{blue}$R$};
\draw (0,1) node {\small $R_1$};
\draw (0,-1) node {\small $R_2$};
\end{tikzpicture} \qquad \qquad
\begin{tikzpicture}[scale=1, baseline={([yshift=0cm]current bounding box.center)}]
\draw[thick] (0,0) circle (1.5);
\draw[thick,blue] (1.3,.7) arc (300:240:2.6);
\draw[thick,blue] (-1.3,-.7) arc (120:60:2.6);
\draw[fill,black] (-1.3,-.68) circle (0.1);
\draw[fill,black] (1.3,-.68) circle (0.1);
\draw[fill,black] (-1.3,0.68) circle (0.1);
\draw[fill,black] (1.3,0.68) circle (0.1);
\draw (0,0) node {\footnotesize  $R_3$};
\draw (0,.5) node {\footnotesize \color{blue}$R_A$};
\draw (0,-.5) node {\footnotesize \color{blue}$R_B$};
\draw (0,1) node {\footnotesize  $R_1$};
\draw (0,-1) node {\footnotesize  $R_2$};
\draw (0,1.88) node {\small \color{red}$m_1$};
\draw (0,-1.85) node {\small \color{red}$m_2$};
\draw (1.85,0) node {\small \color{red}$m_3$};
\draw (-1.85,0) node {\small \color{red}$\tilde{m}_3$};
\draw (1.09,0.35) node {\small \color{red}$m_A$};
\draw (-1,0.35) node {\small \color{red}$\bar{m}_A$};
\draw (1.09,-0.35) node {\small \color{red}$\bar{m}_B$};
\draw (-1,-0.35) node {\small \color{red}$m_B$};
\draw (-1.75,-.75) node {\footnotesize $\tau_3$};
\draw (-1.75,.75) node {\footnotesize $\tau_2$};
\draw (1.75,-.75) node {\footnotesize $\tau_4$};
\draw (1.75,.75) node {\footnotesize $\tau_1$};
\end{tikzpicture}
\end{align}

\noindent A further example is the four-point function with crossed connections into bilocals. Then the lines associated with the bilocals cross in the bulk of the diagram. The expectation value of the product of two bilocal operators $\mathcal{O}_{R_A,m_A\bar{m}_A}(\tau_1,\tau_3)$ and $\mathcal{O}_{R_B,m_B\bar{m}_B}(\tau_2,\tau_4)$, with time-ordering $\tau_1<\tau_2<\tau_3<\tau_4$, is given by 
\begin{align}
\begin{tikzpicture}[scale=1, baseline={([yshift=0cm]current bounding box.center)}]
\draw[thick]  (0,0) ellipse (1.6 and 1.6);
\draw[thick,blue] (1.1,-1.2) arc (35.7955:82:4);
\draw[thick,blue] (-1.1,-1.2) arc (144.2045:98:4);
\draw (0,-0.8) node {\small $R_3$};
\draw (0,1) node {\small $R_1$};
\draw (-1,-0.2) node {\small $R_2$};
\draw (1,-0.2) node {\small $R_4$};
\draw (-0.5,0.35) node {\small \color{blue}$R_A$};
\draw (0.5,0.35) node {\small \color{blue}$R_B$};
\draw[fill,black] (-1.56,0.4) circle (0.1);
\draw[fill,black] (1.56,0.4) circle (0.1);
\draw[fill,black] (-1.11,-1.17) circle (0.1);
\draw[fill,black] (1.11,-1.17) circle (0.1);
\draw (-1.9,0.4) node {\small $\tau_1$};
\draw (-1.45,-1.4) node {\small $\tau_2$};
\draw (1.45,-1.4) node {\small $\tau_3$};
\draw (1.9,0.4) node {\small $\tau_4$};
\draw (0,1.85) node {\small \color{red}$m_1$};
\draw (0,-2) node {\small \color{red}$m_3$};
\draw (1.97,-.75) node {\small \color{red}$m_4$};
\draw (1.11,.6) node {\small \color{red}$\bar{m}_B$};
\draw (-1.11,-0.7) node {\small \color{red}$m_B$};
\draw (-1.1,.6) node {\small \color{red}$m_A$};
\draw (1.11,-0.7) node {\small \color{red}$\bar{m}_A$};
\draw (-1.97,-.75) node {\small \color{red}$m_2$};
\end{tikzpicture} 
\end{align}
Using the diagrammatic rules, one writes:
\begin{align}
\nonumber \Big\langle \mathcal{T} \mathcal{O}_{R_A,m_A\bar{m}_A}&(\tau_1,\tau_3) \mathcal{O}_{R_B,m_B\bar{m}_B}(\tau_2,\tau_4) \Big\rangle =\sum_{R_1,R_2,R_3,R_4}\prod_i \left(\dim R_ie^{-a L_i C_{R_i}}\right)\sj{R_B}{R_1}{R_4}{R_A}{R_3}{R_2}\\
&\cross \sum_{m_1,m_2,m_3,m_4}\tj{R_1}{R_A}{R_2}{m_1}{m_A}{m_2}\tj{R_1}{R_B}{R_4}{m_1}{\bar{m}_B}{m_4}\tj{R_2}{R_B}{R_3}{m_2}{m_B}{m_3}\tj{R_3}{R_A}{R_4}{m_3}{\bar{m}_A}{m_4}.\label{pgOTO4pt}
\end{align}
Such a configuration is closely related to, but not equal to an out-of-time ordered (OTO) correlator.\footnote{Schematically, imagine we can write out the bilocals as the product of local operators. Then what we computed above is
\begin{equation}
\contraction{}{\mathcal{O}_1}{\mathcal{O}_2}{\mathcal{O}_3}
\bcontraction{\mathcal{O}_1}{\mathcal{O}_2}{\mathcal{O}_3}{\mathcal{O}_4}
\left\langle \mathcal{O}_1\mathcal{O}_2\mathcal{O}_3\mathcal{O}_4\right\rangle,
\end{equation}
whereas the genuine OTO-correlator obtained by swapping operators in a time-ordered correlator using the braiding $R$-matrix, would be \\
\begin{equation}
\text{$\contraction[1ex]{}{\mathcal{O}_1}{}{\mathcal{O}_2}
\left\langle \mathcal{O}_1 \mathcal{O}_2 \right.
\hspace{-0.2cm}
\bcontraction[1ex]{}{\mathcal{O}_3}{}{\mathcal{O}_4}
\left.\mathcal{O}_3\mathcal{O}_4\right\rangle$
$\quad \stackrel{\mathclap{\normalfont\mbox{R}}}{\Longrightarrow} \quad $
$\contraction{}{\mathcal{O}_1}{\mathcal{O}_3}{\mathcal{O}_2}
\bcontraction{\mathcal{O}_1}{\mathcal{O}_3}{\mathcal{O}_2}{\mathcal{O}_4}
\left\langle \mathcal{O}_1\mathcal{O}_3\mathcal{O}_2\mathcal{O}_4\right\rangle$}.
\end{equation}
The resulting expressions are very closely related though: putting operators out of time order is effectively forcing the bilocal lines (to be identified as Wilson lines in the bulk theory) to cross, in the end reproducing the same computation. The appearance of the $6j$-symbol for the holographic $SL(2,\mathbb{R})$ BF-theory from crossing Wilson lines is indeed related to the OTO correlation functions of the boundary Schwarzian theory, which is discussed elsewhere \cite{Blommaert:2018.2}. Such an investigation is also being pursued independently \cite{IPW}.}

\subsection{Boundary Anchored Wilson Lines in 2d Yang-Mills}
In the above, the diagrams were merely tools to write down a general amplitude. Our goal now is to demonstrate that the interior of the diagram can be interpreted as the 2d  Yang-Mills bulk, with the internal lines interpretable as bulk Wilson lines: we want to prove here that the above correlators encode 2d YM correlation functions of boundary-anchored Wilson lines in the interior of the disk.

\begin{figure}[h]
\centering
\includegraphics[width=0.8\textwidth]{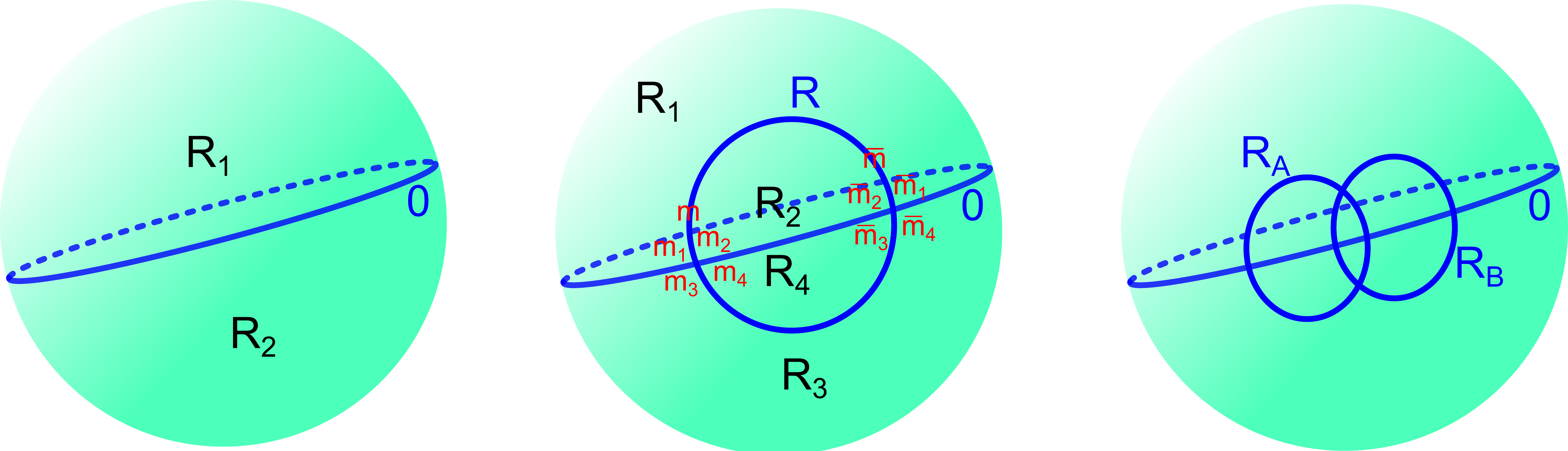}
\caption{Left: cutting a sphere along an identity Wilson loop causes a decomposition into two disks. Middle: cutting a sphere through an additional Wilson loop produces two disks with single boundary-anchored Wilson lines. Right: two additional intersecting Wilson loops leads to two disks that each contain two crossing boundary-anchored Wilson lines.}
\label{Deconstruct1}
\end{figure}

This specific subset of correlation functions has not been deduced in the literature yet, but can be obtained from it by suitable manipulations. In this section, we resort to a deconstructive method by cutting open known YM sphere path integrals into two disks, where the cutting line crosses a suitable number of Wilson loops (Figure \ref{Deconstruct1}).\footnote{A constructive method similar to the original calculation of Wilson line correlators in YM \cite{Witten:1991we} is given in \cite{Blommaert:2018.2}.} This leads to disk correlation functions of boundary-anchored Wilson lines. We heavily draw upon the results of 2d YM, which can e.g. be found in the review \cite{Cordes:1994fc}.
\\~\\
Cutting open the path integral on the sphere along a certain line is achieved naturally by inserting an identity Wilson loop along this cut, as we will show by example. Consider the path integral of YM on a sphere with a Wilson line $\mathcal{W}_{R}$ in irrep $R$ inserted: 
\begin{equation}
\Big\langle \mathcal{W}_{R}\Big\rangle =\sum_{R_1,R_2}\dim R_1 \dim R_2 \, e^{-A_1C_{R_1}}e^{-A_2C_{R_2}}\int d U \chi_{R_1}(U)\chi_{R}(U)\chi_{R_2}(U^{-1}),\label{1wilson}
\end{equation}
where $\chi(R)$ denotes the character in representation $R$ and $U = P \, \exp(\oint A)$ is the holonomy along the spatial slice of interest. The integral over $U$ is readily evaluated explicitly using:
\begin{equation}
\int d U \chi_{R_1}(U)\chi_{R}(U)\chi_{R_2}(U)=N_{R_1RR_2}, \quad \chi_{\bar{R}}(U) = \chi_R(U^{-1})
\end{equation}
Consider now the special case where $R=0$ i.e. the identity representation, depicted in Figure \ref{Deconstruct1} (left). The LHS of \eqref{1wilson} becomes just the sphere partition function $\big\langle 1 \big\rangle=Z$, and \eqref{1wilson} decomposes as:
\begin{equation}
Z=\int d U Z_\text{top}(U) Z_\text{bot}(U^{-1}),
\end{equation}
where $Z_\text{top}(U)$ is the disk amplitude with boundary holonomy $U$ \eqref{2dYMzU}.
\\~\\
A second, more illustrative example is obtained by considering the sphere path integral of two crossing Wilson lines of which one is in the identity representation, as depicted in the middle Figure \ref{Deconstruct1}. The diagrammatic rules for calculating such a diagram in YM are well-known, and were established by Witten \cite{Witten:1991we}. Each of the crossings contains a $6j$-symbol, which arises as summing the product of four $3j$-symbols over all relevant $m_i$ labels, each representing three adjacent irreps. Two of those four $3j$-symbols then contain the identity irrep from the separating Wilson loop. These can be written as:
\begin{align}
\tj{R_1}{0}{R_3}{m_1}{0}{m_3}\tj{R_1}{0}{R_3}{\bar{m}_1}{0}{\bar{m}_3} &= \int d W R_1(W)_{m_1\bar{m}_1}R_3(W)_{m_3\bar{m}_3}, \nonumber \\ \label{gluesegment}
\tj{R_2}{0}{R_4}{m_2}{0}{m_4}\tj{R_2}{0}{R_4}{\bar{m}_2}{0}{\bar{m}_4} &= \int d V R_2(V)_{m_2\bar{m}_2}R_4(V)_{m_4\bar{m}_4}
\end{align}
where we introduce the boundary group elements $W$ and $V$ along the relevant segments. This orthonormality relation of the representation matrices is a special case of the more general formula
\begin{equation}
\int d U \, R_1(U)_{n_1m_1} R_2(U)_{n_2m_2} R_3(U)_{n_3m_3}=\tj{R_1}{R_2}{R_3}{m_1}{m_2}{m_3}\tj{R_1}{R_2}{R_3}{n_1}{n_2}{n_3},
\end{equation}
The sphere amplitude with the Wilson line insertion is found to decompose as:
\begin{equation}
\Big\langle \mathcal{W}_{R}\Big\rangle=\sum_{\bar{m}\bar{n}}\int d V d W Z_\text{top}(V,W)Z_\text{bot}(V,W),\label{1wilsonglue}
\end{equation}
where
\begin{align}
Z_\text{top}(V,W)=\sum_{R_1,R_2}\dim R_1 \dim R_2 \sum_{m_1,\bar{m}_1,m_2,\bar{m}_2}&R(W)_{m_1\bar{m}_1}R(V)_{m_2\bar{m}_2}e^{-A_1 C_{R_1}} e^{-A_2 C_{R_2}} \nonumber \\
&\times\tj{R_1}{R_2}{R}{m_1}{m_2}{m}\tj{R_1}{R_2}{R}{\bar{m}_1}{\bar{m}_2}{\bar{m}}.\label{ztop}
\end{align}
It is not hard to check that performing the integrals over $V$ and $W$ in \eqref{1wilsonglue} results back in \eqref{1wilson}. We identify this as the disk path integral of a boundary-anchored Wilson line, with boundary holonomy $W$ resp. $V$ on the two relevant open boundary intervals. Setting $V=W=1$, to obtain a disk with a boundary, we enforce $m_i = \bar{m}_i$ and obtain \eqref{pg2pt}, with a suitable choice of the parameters $a$ and $L_i$.
\\~\\
As a final example we can consider three crossing Wilson loops (in the sense of Olympic rings, not Audi rings) of which one is in the identity representation (Figure \ref{Deconstruct1} right). To each segment of the identity line we apply \eqref{gluesegment}. The sphere amplitude is observed to decompose as:
\begin{equation}
\Big\langle \mathcal{W}_{R_A}\mathcal{W}_{R_B} \Big\rangle_{\text{crossing}} = \sum_{m_A,\bar{m}_A,m_B,\bar{m}_B} \left(\prod_i \int d V_i\right) Z_\text{top}(V_j) Z_\text{bot}(V_j),
\end{equation}
where there are now four distinct integration variables $V_1,V_2,V_3,V_4$. 
The disk partition function is given by a lengthy expression similar to \eqref{ztop}, now with a product of 4 representation matrices, 4 3j symbols and a 6j symbol. Setting the holonomies equal to 1: $V_i=1$, we reproduce \eqref{pgOTO4pt} with a suitable choice of $a$ and $L_i$.

\subsection{Wilson Lines as Boundary Bilocals}
The purpose of this section is to make explicit to which extent there is a 1-to-1 mapping of particle-on-a-group correlators to 2d Yang-Mills correlators. Correlators for the particle-on-a-group model are characterized by time differences between operator insertions. Two-dimensional Yang-Mills theory on the other hand would claim these time diferences are unphysical: they are associated to parts of the area-preserving diffeomorphism and are hence gauge-variant: only areas have meaning in 2d YM.

There is though, a 1-to-1 mapping from the parameters $L_i$ labeling correlators on the boundary theory, to parameters of bulk correlators (the areas $A_i$): $a L_i = A_i$. In light of the above observations, this formula immediately implies a direct identification between correlators in both theories. 

\begin{figure}[h]
\centering
\includegraphics[width=0.8\textwidth]{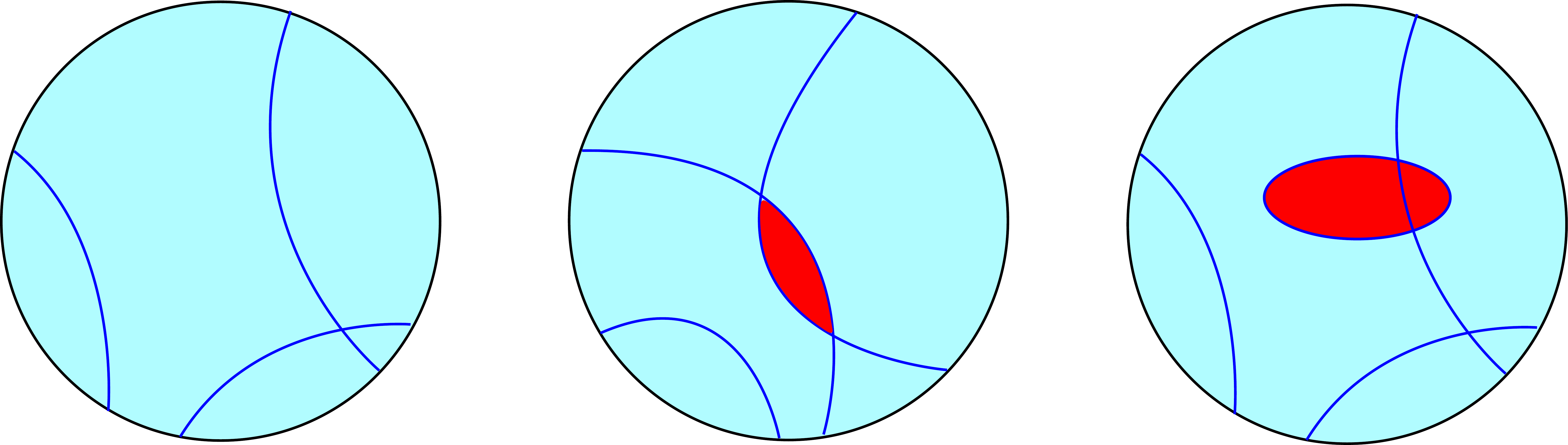}
\caption{Three examples of a Wilson line network in 2d YM. The red regions cannot be reproduced from the edge theory perspective.}
\label{classWilson}
\end{figure}

\begin{figure}[h]
\centering
\includegraphics[width=0.95\textwidth]{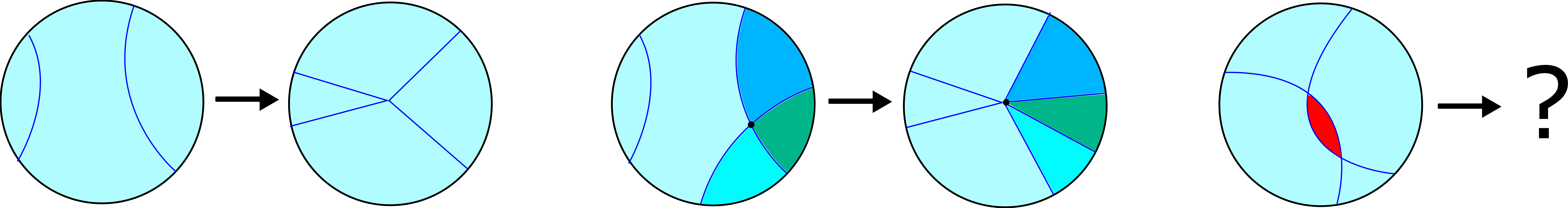}
\caption{Area-preserving deformation of Wilson line network into disk wedges. This is only possible if there is no enclosed area in the bulk.}
\label{WilsonRadial}
\end{figure}

Not every bulk correlator is contained within our edge theory though (Figure \ref{classWilson}). The red regions in Figure \ref{classWilson} are examples of bulk-boundary interactions, that are not captured by the edge action. An example of bulk-bulk interactions is a configuration of Wilson loops that doesn't reach the boundary. As mentioned at the beginning of section \ref{s:ymba} and made explicit here, the boundary action \emph{by construction} does not capture these interactions.
This perspective neatly interpolates between fully topological theories (such as 2d BF theory, which is entirely holographically dual to the particle-on-a-group model) where Wilson lines can be deformed entirely to the boundary and the edge theory contains everything, and non-topological theories (such as $d>2$ YM) where the edge theory only contains the punctures of the Wilson lines with the boundary \cite{Blommaert:2018rsf}. For the quasi-topological 2d YM theory, the edge theory does reproduce many aspects of the bulk, but not everything.
\\~\\
There is an interesting explicit way of visualizing the mapping of only this class of Wilson lines to the boundary correlation functions. Consider performing an area-preserving diffeomorphism to deform the bulk Wilson line network into one that only contains wedges of the disk (Figure \ref{WilsonRadial}). Such a procedure is only possible when there is no area contained fully within the bulk.

Imagine now we insert such a boundary-anchored Wilson line in irrep $R$ into our edge computation of section \ref{sect:YM}. We need to perform an on-shell evaluation of the Wilson line:
\begin{equation}
\mathcal{W}_{R}(\tau_i,\tau_f)=\mathcal{P}\exp{\int_\mathcal{C}A}=R(g(\tau_f))\mathcal{P}\exp{\int_\mathcal{C}A(\mq)}R(g^{-1}(\tau_i)),
\end{equation}
where $A= gA(\mq)g^{-1}-dgg^{-1}$ and $A(\mq)$ is the particular solution of section \ref{s:2dym}. The Wilson lines in Figure \ref{WilsonRadial} runs along constant $\tau$ lines everywhere, except for a sharp turn near $\rho=0$. Because of our (residual) gauge choice $A_\rho (\mq)=0$ in section \ref{s:2dym}, the Wilson line is just:
\begin{equation}
{\mathcal{W}_{R}(\tau_i,\tau_f)}_{m\bar{m}}=R(g(\tau_f)g^{-1}(\tau_i))_{m\bar{m}},
\end{equation}
and, crucially, does not influence the path integral over the gauge-fixed field $A_\mu$. This formula is precisely the definition of the bilocal operators \eqref{bilocal} we identified earlier from the 2d WZW perspective. These wedge-diagrams make explicit the mapping of particle-on-a-group correlators to (area-preserving equivalence classes of) boundary anchored Wilson lines in YM which completes the proof.

One should not attach too much significance to this precise geometric form of the wedge Wilson lines: their apparent importance hinges on the gauge choice $A_\rho=0$ and the area-preserving diffeomorphism of Figure \ref{WilsonRadial} is tailored to this gauge choice. The important part is that, within this gauge choice, we understand why the class of boundary-anchored Wilson lines without bulk area enclosures are being computed by boundary bilocal correlators.
\\~\\
This entire construction can of course also be done for Maxwell. This is an instructive exercise, and we refer the reader to appendix \ref{sect:u1} for some details.

\subsection{The Horizon Theory as Topological Quantum Mechanics}
We mentioned earlier that the edge theories on the Rindler horizon can be viewed as arising from decoupled 2d Maxwell and YM edge theories: this statement was made explicitly in section \ref{s:YMrindler} where we noticed that the edge theory becomes ultralocal in the transverse $\mathbf{x}$-directions. The path integral was that of \eqref{zsinftyym}:
\begin{equation}
\label{zsinftyym2}
Z=\int \left[\dpi g\right]\left[\dpi \mq\right]\exp{i\int d\tau d\mathbf{x}\Tr \left(\mq\partial_\tau g g^{-1}\right)}.
\end{equation}
Correlators of this Rindler edge theory can be obtained directly from the 2d results described above. In particular, the decoupling argument implies that correlators of bilocals at distinct $\mathbf{x}$-points factorize:
\begin{equation}
\left\langle R(g(\tau_2,\mathbf{x}_2))_{mn}R(g^{-1}(\tau_1,\mathbf{x}_1))_{n\bar{m}}\right\rangle = \left\langle R(g(\tau_2,\mathbf{x}_2))_{mn}\right\rangle\left\langle R(g^{-1}(\tau_1,\mathbf{x}_1))_{n\bar{m}}\right\rangle,
\end{equation}
where both expectation values are generally non-zero and could in principle be computed within the particle-on-a-group model. This requires knowledge of local, but gauge-variant correlation functions.\footnote{For the Schwarzian theory, which is the irrational cousin of the particle-on-a-group model, such correlators were computed using the Knizhnik-Zamolodchikov equations in \cite{Mertens:2017mtv}.} \\
A bilocal expectation value at the same spatial point $\mathbf{x}$ can be computed and is given by:
\begin{align}
\left\langle \mathcal{O}_{R,m\bar{m}}(\tau_1,\tau_2,\mathbf{x}) \right\rangle = \sum_{R_1,R_2}\dim R_1\dim R_2 \sum_{m_1,m_2}\tj{R_1}{R}{R_2}{m_1}{m}{m_2}\tj{R_1}{R}{R_2}{m_1}{\bar{m}}{m_2} = \left\langle 1\right\rangle \delta_{m\bar{m}},
\end{align}
which can be checked explicitly by using $\sum_{R_2} N_{R_1RR_2}\text{dim }R_2  = \text{dim }R_1 \text{ dim }R$. As the bilocal operator can be neutralized by taking $\tau_2 \to \tau_1$ (the correlator is independent of the time parameter $\tau_{21}$), any uncrossed correlator evaluates to just the partition function $Z=\left\langle 1\right\rangle$ itself, which is manifestly UV-divergent. Ultimately, the divergence arises from the infinite dimensionality of the 1d particle-on-a-group Hilbert space. The only non-trivial correlators to be computed are crossed correlators, where the crossed operator ordering leads as before to the $6j$-symbol as the only non-trivial feature. 
\\~\\
The resulting edge theory, while severely UV-divergent, can be viewed as topological quantum mechanics.\footnote{Similar theories were studied in different contexts in \cite{Witten:2010zr,Dedushenko:2016jxl,Mezei:2017kmw}.} In 1d, this also means the edge theory is conformally invariant. This is due to the well-known dimensional argument that a 1d CFT has the density of states $\rho(E) = A\delta(E) + B/E$, where $B=0$ to have any meaningful low-energy theory. So $\rho(E) \sim \delta(E)$ i.e. a theory of ground states. The Hamiltonian is identically zero $H \equiv 0$, which can be seen as well by noting that \eqref{zsinftyym2} is time-reparametrization invariant. Its excitations contain no energy, and correlators do not depend on time differences, but only on the ordering of the operators. The horizon correlators can be obtained by taking the $e\to0$ limit of 2d YM. Introducing the coupling constant $e$ in the 2d Yang-Mills action, and introducing a Lagrange multiplier $\psi$, it can be rewritten as \cite{Witten:1992xu}:
\begin{equation}
-\frac{1}{2e}\int d^2x\sqrt{-g}\, \Tr F^2\to \frac{e}{2} \int d^2 x\sqrt{-g}\, \Tr \psi^2 + \int \Tr \psi F.
\end{equation}
In the limit $e\to 0$ the above action reduces to just
\begin{equation}
S\left[A,\psi\right]=\int \Tr \psi F,
\end{equation}
which is the topological BF theory, in this case without boundary dynamics.
\\~\\
A somewhat tantalizing but highly speculative idea might be that a UV-complete theory (such as string theory), could lead to a replacement of the group $G$ with its quantum extension $G_q$. The latter is known to include only a finite number of irreducible representations, and hence allows the possibility of a UV-complete horizon theory. This is indeed a known strategy to get rid of divergences within the spinfoam formulation of loop quantum gravity \cite{Turaev:1992hq,Crane:1994ji}.

\section{Discussion}
The centerpiece of this work has been the construction of a boundary action describing edge mode dynamics in Maxwell and non-Abelian Yang-Mills theory. One obtains this boundary action by explicitly sourcing the Yang-Mills theory on the boundary, and then path integrating out the bulk degrees of freedom. The resulting boundary action is the on-shell evaluated sourced action. The boundary degrees of freedom are the current (the source) and the large would-be gauge degrees of freedom which have become dynamical fields. 

This procedure is rigorous for the free abelian case. For the non-abelian case (with a quartic action) we \emph{choose} to focus on the boundary action only and neglect bulk-boundary and bulk-bulk interactions. Though this does not capture the full Yang-Mills theory, we claim that it is a sensible procedure. As evidence of this, the correct boundary commutators were obtained from our edge action. Also, our analysis in 2d led to a full solution of this question: one can clearly identify which part of the full Yang-Mills theory is captured by the boundary theory, and which part is not. This has been discussed in section \ref{s:cor}. 

The edge action we constructed \eqref{action} is in general a higher-derivative theory, and we gave a preliminary analysis of these specific theories in appendix \ref{app:constraint}. It would be useful to obtain a more complete understanding of the canonical structure, an analysis that we postpone to possible future work.
\\~\\
The procedure is in many ways identical to how boundary actions arise in topological gauge theories such as 3d Chern-Simons \cite{Elitzur:1989nr,Carlip:1991zm} and 2d BF \cite{Mertens:2018fds,Gonzalez:2018enk}, and gravitational theories such as AdS$_3$ gravity \cite{Coussaert:1995zp, Carlip:2005tz} and 2d Jackiw-Teitelboim \cite{Jensen:2016pah,Maldacena:2016upp,Engelsoy:2016xyb} and its flat limit \cite{Verlinde:1991iu,Dubovsky:2017cnj}. 

For example, one obtains the 2d WZW action from 3d Chern-Simons by evaluating the Chern-Simons action on-shell after imposing appropriate boundary conditions. This on-shell evaluation stems from a path integral over the bulk degrees of freedom; in this topological example the determinant of quadratic fluctuations is simply the identity and one is left directly with the WZW path integral. Large gauge fields in Chern-Simons are the physical degrees of freedom of the boundary WZW theory. Within Chern-Simons, the boundary edge perspective has already been useful to think about entanglement entropy (see e.g. \cite{Qi:2012,Geiller:2017xad,Fliss:2017wop,Wong:2017pdm}).

There is though, at least one difference with the procedure to obtain the Yang-Mills boundary action: the origin of the boundary action is different. For Yang-Mills, the boundary $\mj A$-coupling term is manifestly gauge-variant and splits up as 
\begin{equation}
\mj A = \mj gA[\mj]g^{-1} - \mj dg g^{-1}.
\end{equation}
Within Chern-Simons, introducing a boundary current as in \eqref{Gs}, the boundary conditions become: $\mj^\mu=\left.\epsilon^{\mu\nu\sigma}n_\nu A_\sigma\right|_\text{bdy}$, such that the on-shell evaluation of the boundary contribution to the action vanishes: 
\begin{equation}
\Tr(\mj A)=n_\nu \sum_a \epsilon^{\mu\nu\sigma}A_\sigma^a A_\mu^a=0.
\end{equation}
The bulk CS action is explicitly gauge-variant, and the WZW action stems completely from the on-shell evaluation of this bulk action.
\\~\\
Several applications of the boundary action \eqref{action} have been discussed. These serve as consistency checks, but are also interesting in their own rights.
\\~\\
From the boundary path integral for Maxwell in Rindler we recover the full Maxwell thermal partition function including the Kabat contact term. The edge states are identified as electric charges on the boundary, or (dual) large gauge degrees of freedom on the boundary. Infinite redshift leads to a localization of the boundary path integral, proving the absence of magnetostatic edge states. This calculation embeds the discussion of \cite{Blommaert:2018rsf} in a more generic context. In \cite{Blommaert:2018rsf}, we recovered the Maxwell edge partition function by directly quantizing the static sector of Maxwell theory; which supplies the calculation of Donnelly and Wall \cite{Donnelly:2012st} with an underlying canonical structure. The canonical algebra of the edge degrees of freedom was inferred from the usual Maxwell canonical structure, identifying Wilson line punctures on the boundary (large gauge transformations) as canonical conjugates to electric flux through the boundary. The same interpretation follows from the path-integral perspective of this work.

In a different context \cite{Strominger:2017zoo,Hawking:2016msc,Hawking:2016sgy}, Strominger et al. identified configurations labeled by different large gauge configurations, or equivalently by boundary charge configurations, at asymptotic null infinity as inequivalent vacua associated with different soft photon configurations. The identification of these soft photons as the Maxwell edge modes has been highlighted in \cite{Blommaert:2018rsf}. It would be interesting to perform the computation of this work directly at future and past null infinity, generalizing it from a spacelike surface and a null horizon surface.

Our generalization of this computation to Yang-Mills on the Rindler horizon in section \ref{s:YMrindler} is new, and provided us with an additional motivation: the edge theory of any horizon is directly related to a 2d Yang-Mills computation.

The edge action on the Rindler horizon \eqref{zsinftyym2} was identified as a version of topological quantum mechanics, and it would be interesting to pursue this line of thought further.
\\~\\
As a second application, we discussed 2d Yang-Mills on a disk: a quasi-topological theory that has been completely solved \cite{Witten:1991we}. The boundary edge theory was shown to be a particle on a group whose worldline is the boundary of the disk. It is in this specific example that our boundary action shows its true colors. We go far beyond calculating the partition function, and identify correlators of boundary-anchored Wilson lines in 2d Yang-Mills as correlators of bilocal operators in the particle-on-a-group model. For $d>2$, we expect the edge theory to contain only information of the Wilson line punctures. For a topological theory, the entire Wilson line is contained within its endpoints only and the edge theory becomes the entire theory. The 2d Yang-Mills example is a non-trivial intermediate case where the edge theory contains more than just the boundary surface, but it does not carry all information about the bulk.
\\~\\
There is reason to assume that the range of application of the boundary path integral of this work extends beyond spin-1 gauge theories. There is at least one particular example we know of where this line of reasoning works. Repeating the construction of section \ref{s:rindler} in Rindler space for linearized gravity - where the boundary current is a rank-2 tensor $T$ representing energy-momentum in the boundary - we show elsewhere \cite{Blommaert:2018} that one obtains the analogue of formulas \eqref{mwcountingcharges} and \eqref{mwcountinggauge}:
\begin{equation}
Z^\text{bdy}=\int \left(\prod_{j}\left[\dpi \mathcal{P}_j\right]\right)\delta\left(\partial^i \mathcal{P}_i\right)=\int \left(\prod_{j}\left[\dpi A_j\right]\right)\delta\left(\partial^i A_i\right),
\end{equation}
where $\mathcal{P}_j$ are just the spatial charges associated with $T$, generating diffeomorphisms in the boundary surface. The large diffeomorphisms consist of a massless divergence-free vector field $A$. As before, we expect the Rindler edge action to be described by a topological field theory; perhaps the irrational analogue of the $e\to0$ limit of the particle-on-a-group model, which is Schwarzian QM with Lagrangian $L \sim \frac{1}{G_N} \int d\mathbf{x} \left\{f(\mathbf{x},\tau),\tau\right\}$ where $G_N \to 0$ to obtain the topological version. It would be interesting to understand this.

In \cite{Blommaert:2018} we will show that this boundary contribution amounts to the correct graviton partition function, including the Kabat contact term, a result that can be checked directly by identifying the contribution of the graviton to the bosonic string partition function.
\\~\\
It should be possible to apply this construction directly to the action of open string field theory. It is natural to expect that the large BRST exact fields will become dynamical boundary fields. That these large gauge fields contain the boundary degrees of freedom was recently suggested in \cite{Balasubramanian:2018axm} from the perspective of the presymplectic potential. We hope to be more specific about these stringy boundary degrees of freedom in \cite{Blommaert:2018}.
It would in any case be interesting to understand better the role of edge states within string theory and its relation to the old Susskind-Uglum picture of horizon-piercing strings (see e.g. \cite{Donnelly:2016jet} for a 2d example, and \cite{Susskind:1994sm,He:2014gva,Mertens:2015adr,Mertens:2016tqv} for a variety of results in this direction using Euclidean techniques).

\section*{Acknowledgements}
The authors thank Nele Callebaut, Marc Geiller, Jutho Haegeman, Luca Iliesiu, Ho Tat Lam, Gustavo Turiaci, Dieter Van den Bleeken, Laurens Vanderstraeten, Bram Vanhecke and Herman Verlinde for several valuable discussions. AB and TM gratefully acknowledge financial support from Research Foundation Flanders (FWO Vlaanderen).

%%%%%%%%%%%%%%%%
% APPENDICES %
%%%%%%%%%%%%%%%%
\appendix

\section{Gluing}
\label{sect:gluing}
The question arises on how to recover the partition function of a theory on $\mathcal{M}\cup \bar{\mathcal{M}}$ from the separate theory on $\mathcal{M}$ and $\bar{\mathcal{M}}$. 

\subsection{Gauge Bundle View on Gluing}
\label{sect:fiber}
As an alternative to the procedure sketched in section \ref{sect:Maxwell}, we can glue the two halves together using gauge-fixed fields $A$ and $\bar{A}$ instead. Then the gluing only requires these fields to be equal up to a gauge transformation: \eqref{mwdeltapij} is replaced by
\begin{equation}
\frac{1}{\text{vol} G_\partial}\int \frac{\left[\dpi\psi\right]}{\text{vol} G_\partial}\left[\mathcal{D}\mathcal{J}^\mu\right] e^{i\int_{\partial \mathcal{M}}d^{d-1}x\, \mj ^\mu ( A_\mu-\bar{A}_\mu + \partial_\mu \psi)},\label{bundleglue}
\end{equation}
where analogous to \eqref{volgd}
\begin{equation}
\text{vol} G_\partial=\int \left[\dpi\psi\right]=\delta(0). 
\end{equation}
In \eqref{bundleglue} it is understood that $A$ and $\bar{A}$ are the gauge-fixed fields as in e.g. \eqref{mwactiontotal}. This gluing is along the lines of the construction of a gauge/fiber bundle, where the two halves correspond to two patches with the boundary the common region. The compatibility condition of fiber bundle theory then indeed requires the fields $A$ and $\bar{A}$ on the boundary to be linked by a gauge transformation.

After on-shell evaluation, \eqref{bundleglue} can be expanded, and one obtains the gluing formula:
\begin{equation}
\int \left[\dpi \mj^\alpha\right]\left[\dpi \bar{\mj}^\alpha\right]\left[\dpi \phi\right]\left[\dpi \bar{\phi}\right]\delta(\mj+\bar{\mj}) \delta(\phi+\bar{\phi}) e^{-S\left[\mj,\phi\right]} e^{-S\left[\bar{\mj},\bar{\phi}\right]},
\end{equation}
where distinct currents and would-be gauge fields have been introduced for the left and right regions and where $S\left[\mj,\phi\right]$ is the thermal boundary action \eqref{mwsEuclidean}:
\begin{equation}
S^\text{bdy}\left[\mj,\phi\right]=-\int_0^\beta d\tau \int d^{d-2}x\left(\frac{1}{2}\mj^\alpha A[\mj]_\alpha + \mj^\alpha \partial_\alpha\phi\right).
\end{equation}
From the gluing perspective, the only physical combination of $\phi$ and $\bar{\phi}$ is $\psi$. The other independent combination of $\phi$ and $\bar{\phi}$ is unphysical, and is removed by the $\delta(\phi+\bar{\phi})$ delta-functional.
\\~\\
Integrating out $\phi$ and $\bar{\phi}$ and path integrating over $\bar{\mj}$ to enforce the gluing constraint, one obtains
\begin{equation}
Z=\int \left[\dpi \mj^\alpha\right]\delta (\partial_\alpha \mj^\alpha) \exp{\int_0^\beta d\tau\int d^{d-2}x \frac{1}{2}\mj^\alpha \left(A(\mj)_\alpha-\bar{A}(-\mj)_\alpha\right)}.\label{gluingpi1}
\end{equation}
Assuming the action is quadratic in $\mj$, one can alternatively write the gluing fully in terms of the large gauge transformations as:
\begin{equation}
Z=\int \left[\dpi \phi_L\right]\left[\dpi \phi_R\right]\delta(\mj(\phi_L)+\bar{\mj}(\phi_R)) \exp{-S^{\text{bdy}}[\phi_L]-S^\text{bdy}[\phi_R]},\label{gluingpi2}
\end{equation}
in terms of the individual actions $S^{\text{bdy}}[\phi]$ \eqref{mwgaugeaction} for each side. An explicit example provides further clarification.

\subsection{Example I. 2d Maxwell}
As an explicit example we investigate 2d Maxwell, already discussed in section \ref{s:2dmw}. 
The currents reduce to charges $\mq$. Gluing two disks of areas $A$ and $B$ together, following \eqref{gluingpi1}, one obtains:
\begin{align}
\nonumber Z&=\int d\mq d\bar{\mq}\delta (\mq+\bar{\mq})\exp{-A\frac{C(\mq)}{2}}\exp{-B\frac{C(\bar{\mq})}{2}}\\
&=\int d\mq \exp{-(A+B)\frac{C(\mq)}{2}},\label{2dmwsphere}
\end{align}
which is just the partition function of 2d Maxwell on a sphere of total area $A+B$.

In our language, gluing the two partition functions together can also be written as
\begin{align}
\nonumber Z=\int &\left[\dpi \mq \right]\left[\dpi \bar{\mq} \right]\left[\dpi \phi \right]\left[\dpi \bar{\phi} \right]\delta(\mq+\bar{\mq})\delta(\phi+\bar{\phi}) \\
&\cross \exp{-\int_0^\beta d\tau \left(\frac{a}{2}\mq^2-i\mq \partial_\tau \phi\right)}\exp{ -\int_0^\beta d\tau \left(\frac{b}{2}\bar{\mq}^2-i\bar{\mq} \partial_\tau \bar{\phi}\right)}.\label{z}
\end{align}
Path integrating over $\bar{\mq}$ and introducing a new field
\begin{equation}
\psi=\phi-\bar{\phi},\label{psi}
\end{equation}
the path integral reduces to just
\begin{equation}
Z=\int \left[\dpi \mq\right]\left[\dpi \psi\right]\exp{-\int_0^\beta d\tau \left(\frac{a+b}{2}\mq^2-i\mq \partial_\tau \psi\right)},\label{zqpsi}
\end{equation}
Integrating out $\psi$ in \eqref{zqpsi} on obtains the partition function of 2d Maxwell on a sphere \eqref{2dmwsphere}. On the other hand, integrating out $\mq$, one obtains a quadratic boundary action for $\psi$ that is a particle on $U(1)$ with coupling $a+b$:
\begin{equation}
Z=\int \left[\dpi \psi\right]\exp {-\frac{1}{2}\frac{1}{a+b}\int d \tau (\partial_\tau \psi)^2}.
\end{equation}
The latter can be rewritten as
\begin{equation}
Z= \int \left[\dpi \phi_L\right]\left[\dpi \phi_R\right]\delta\left(\frac{1}{a}\partial_\tau \phi_L+\frac{1}{b}\partial_\tau \phi_R\right)\exp {-\frac{1}{2a}\int d \tau(\partial_\tau\phi_L)^2}\exp {-\frac{1}{2b}\int d \tau(\partial_\tau\phi_R)^2},\label{mwzgluephi}
\end{equation}
realizing equation \eqref{gluingpi2} in this explicit example.

\subsection{Example II. 2d Yang-Mills}
\label{app:YM}
2d Yang-Mills theory has two conjugate basis of the Hilbert space: the holonomy basis $\left|U\right\rangle$ and the representation basis $\left|R\right\rangle$. These two bases correspond to our two possible perspectives on the evaluation of the boundary action as presented in section \ref{sect:YM}.

Sewing two disks together in 2d YM is achieved by taking the trace in the holonomy basis \cite{Cordes:1994fc}: 
\begin{equation}
Z=\int d U Z(U)\bar{Z}(U^{-1}),\label{glueU}
\end{equation}
where $Z(U)$ is \eqref{2dYMzU} the partition function on the first disk with boundary state $U$, and $\bar{Z}(U^{-1})$ is the partition function on the second disk. Equivalently we can omit the holonomy basis and glue directly in the irrep basis. One writes 
\begin{equation}
Z(U)=\sum_R \chi_R(U) Z(R),
\end{equation}
as in \eqref{2dYMzU}. The characters $\chi_R(U)=\Tr_R(U)=\bra{U}\ket{R}$ are just the Fourier expansion coefficients transforming between the irrep basis and the holonomy basis. They are orthogonal:
\begin{equation}
\int d U \chi_R(U)\chi_{R'}(U^{-1})=\delta_{RR'}.
\end{equation}
Therefore by applying Parseval's theorem for Fourier transforms, we can rewrite \eqref{glueU} and glue directly in the irrep basis:
\begin{equation}
Z=\sum_R Z(R) \bar{Z}(R).
\end{equation}
Explicitly, gluing a disk of area $A$ to a second disk of area $B$, one gets:
\begin{equation}
Z=\sum_{R}(\dim R) ^2\exp{-(A+B)\frac{C(R)}{2}},
\end{equation}
the partition function of 2d Yang-Mills on a sphere of total area $A+B$.

\section{Canonical Structure and Constrained Quantization}
\label{app:constraint}
In this appendix, we initiate a canonical treatment of the boundary action. The analysis in general will be quite complicated due to the non-locality inherent in the boundary action, and we postpone a more elaborate treatment to possible future work. We only focus on the Maxwell action and leave the Yang-Mills analysis as an exercise for the reader. \\
Consider the Maxwell boundary action:
\begin{equation}
S^\text{bdy}\left[\mj,\phi\right]=\int d^{d-1}x\, \left(\frac{1}{2}\mj^\alpha A[\mj]_\alpha + \mj^\alpha \partial_\alpha\phi\right).
\end{equation}
The on-shell evaluation $A[\mj]$ is linear for Maxwell, and we can write it generally as $A_\alpha = Q_{\alpha\beta}J^\beta$ for some operator $Q_{\alpha\beta}$. As the classical sourced solution is found by inverting a differential operator, $Q_{\alpha\beta}$ is generally the inverse of some (possibly non-local) linear differential operator. E.g. in flat space it is given by \eqref{invopflat}. In Rindler space, it is given by \eqref{jat}. So let's write $Q_{\alpha\beta} = \frac{1}{\mathcal{O}}_{\alpha\beta}$. Then we can write
\begin{equation}
\frac{1}{\mathcal{O}}_{\alpha\beta} J^\beta := \chi_\alpha \quad \Rightarrow \quad \mathcal{O}^{\alpha\beta} \chi_\alpha = J^{\beta} \quad \Rightarrow \quad \chi_\alpha = \int dy G_{\alpha\beta}(y,x) J^{\beta},
\end{equation}
in terms of the Green function $G(y,x)$ of the differential operator, that can in principle be computed directly for any given situation. So
\begin{equation}
\mj^\alpha A[\mj]_\alpha = \int dy \mj^\alpha(x) G_{\alpha\beta}(y,x) \mj^\beta(y)
\end{equation}
The above integral contains generally also time $y^0$, making the resulting action highly non-local in space as well as in time. This makes the canonical interpretation much more subtle.
\\~\\
We proceed as follows. Taylor-expanding $Q_{\alpha\beta}$ as a series in higher powers of derivative operators, and only tracking the temporal derivatives, we can view this as a higher-derivative field theory. This can be put in the standard canonical framework by identifying each derivative as a new canonical variable, e.g. $\ddot{q} = q_2$ etc, each with their own conjugate momentum. The resulting system is highly constrained with relations of the type $p_i \sim q_{i+1}$. \\
There is one simple aspect of this system: the $\phi$-field does not figure in the non-local term, and as such we have $\pi_\phi = \mj^0$, irrespective of the non-locality of the remainder. \\
This justifies our statement made in the main text around equation \eqref{mwcommQg}. 
\\~\\
The equations of motion associated to the above system are:
\begin{align}
\delta \phi \quad &\Rightarrow \quad \partial_\alpha \mj^\alpha = 0 \label{eom1} \\
\delta \mj^\alpha \quad &\Rightarrow \quad \partial_\alpha \phi(x) + \int dy G_{\alpha\beta}(y,x) J^{\beta}(y) = 0\label{eom2}
\end{align}
To further deal with this system, we will follow two perpendicular lines of thought. In subsection \ref{s:tikernel}, we assume the operator $\mathcal{O}^{\alpha\beta}$ does not contain time-derivatives. This simplifies things enormously and just requires the analysis of a spatially non-local theory, which can be done by standard techniques. \\
In subsection \ref{s:flkernel}, we discuss the flat space example of section \ref{s:warmup}, for which the bracket 
\begin{equation}
\left\{\phi(\mathbf{x},t),\mj^0(\mathbf{y},t)\right\}=\delta(\mathbf{x}-\mathbf{y})
\end{equation}
and the equations of motion are sufficient to allow a full construction of the Hilbert space.

\subsection{Time-independent Kernel $G_{\alpha\beta}$}
\label{s:tikernel}
Assuming the operator $\mathcal{O}^{\alpha\beta}$ contains no time-derivatives, $G$ only depends on the spatial coordinates $\mathbf{x}$ and $\mathbf{y}$. The Hamiltonian density can be computed using standard techniques, and is given by 
\begin{equation}
\mathcal{H} = - \mj^i \partial_i \phi - \frac{1}{2} J^\alpha \frac{1}{\mathcal{O}}_{\alpha\beta} J^{\beta} = - \mj^i \partial_i \phi - \int d\mathbf{y} \, \frac{1}{2} J^\alpha(\mathbf{x},t) G_{\alpha \beta}(\mathbf{y},\mathbf{x}) J^{\beta}(\mathbf{y},t),
\end{equation}
with canonical momenta:
\begin{equation}
\label{primcons}
\pi_{\phi} = \mj^0, \quad \boxed{\pi_{\mj^i} = 0}.
\end{equation}
The last set of \eqref{primcons} are primary constraints on the system. The total Hamiltonian density (in Dirac's language) is given by
\begin{equation}
\mathcal{H}_T = - \mj^i \partial_i \phi - \frac{1}{2} J^\alpha \frac{1}{\mathcal{O}}_{\alpha\beta} J^{\beta} + \lambda^i  \pi_{\mj^i},
\end{equation}
for multipliers $\lambda^i$ that are determined by Dirac's formalism. The constraints \eqref{primcons} lead to the secondary constraints:
\begin{equation}
\label{seccons}
\dot{\pi_{\mj^i}} \equiv \chi_2 \sim \boxed{\partial_i \phi + \int d\mathbf{y} \, \mj^\alpha(\mathbf{y},t) G_{\alpha i}(\mathbf{y},\mathbf{x}) \approx 0},
\end{equation}
which are just the $\mj^i$ equations of motion \eqref{eom2} themselves. The tertiary constraint leads to a determination of $\lambda^j$:
\begin{equation}
\dot{\chi_2} \sim -\int d\mathbf{y}\, \mj^\alpha(\mathbf{y},t) \partial_{i\mathbf{x}}G_{\alpha 0}(\mathbf{y},\mathbf{x}) + \lambda^j \int d\mathbf{y}\, G_{ji}(\mathbf{y},\mathbf{x}) \approx 0,
\end{equation}
so 
\begin{equation}
\lambda^j = \int d\mathbf{x} \, \mathcal{O}^{ji}\int d\mathbf{y} \mj^\alpha(\mathbf{y},t) \partial_{i\mathbf{x}}G_{\alpha 0}(\mathbf{y},\mathbf{x}).
\end{equation}
This ends the procedure, and we have two constraints \eqref{primcons} and \eqref{seccons}. \\
Using $\left[\pi_{\mj^i}(\mathbf{x},t), \int d\mathbf{z}\, \mj^\alpha(\mathbf{z},t) G_{\alpha j}(\mathbf{z},\mathbf{y})\right] = G_{ij}(\mathbf{x},\mathbf{y})$, the Dirac matrix can now be computed as a $2\times 2$ block matrix:
\begin{equation}
C_{ab}(\mathbf{x},\mathbf{y}) = \left[\begin{array}{cc}
0 & G_{ij}(\mathbf{x},\mathbf{y}) \\
-G_{ij}(\mathbf{x},\mathbf{y}) & 0
\end{array}\right], \quad C^{ab}(\mathbf{x},\mathbf{y}) = \left[\begin{array}{cc}
0 & -\mathcal{O}_{ij}(\mathbf{x},\mathbf{y}) \\
\mathcal{O}_{ij}(\mathbf{x},\mathbf{y}) & 0
\end{array}\right],
\end{equation}
where the inverse is computed using $\int d\mathbf{y} \, C_{ab}(\mathbf{x},\mathbf{y})C^{bc}(\mathbf{y},\mathbf{z}) = \delta_{ac}\delta(\mathbf{x}-\mathbf{z})$. The two constraints are hence second-class. The resulting Dirac brackets can now be computed:
\begin{align}
\left\{\phi(\mathbf{x},t),\pi_\phi(\mathbf{y},t)\right\}_D &= \delta(\mathbf{x}-\mathbf{y}), \\
\left\{\mj^i(\mathbf{x},t), \pi_{\mj^i}(\mathbf{y},t)\right\}_D &= 0, \\
\left\{\mj^i(\mathbf{x},t), \pi_\phi(\mathbf{y},t)\right\}_D &= - \mathcal{O}^{ij}(\mathbf{x},\mathbf{y}) \partial_{j\mathbf{x}} \delta(\mathbf{x}-\mathbf{y}),
\end{align}
and all other brackets vanishing. The last bracket above indeed corresponds to just substituting the eom \eqref{eom2} in the first bracket. These can be used to quantize the system. \\
The $\mj^i$ variables are not independent observables, and only the $\phi$ and $\pi_\phi \equiv \mj^0$ will become the fundamental quantum operators.
\\~\\
One way of writing the Hilbert space is the set of all spatial large gauge transformations $\left|\phi(\mathbf{x})\right\rangle$ or in terms of its conjugate $\left|\mj^0(\mathbf{x})\right\rangle$, the boundary charge distribution. Fourier expanding both fields, the canonical algebra is written as
\begin{equation}
\left[\phi_\mathbf{k},\mj^0_\mathbf{-\mathbf{k}'}\right] = i\delta_{\mathbf{k}\mathbf{k}'},
\end{equation}
which is the structure we found in \cite{Blommaert:2018rsf} for Rindler space. \\
In this case, the thermal path integral computed in \eqref{thpiri} can be directly read as a thermal trace. This is generally so in the case that only static field configurations on the thermal manifold contribute (e.g. due to infinite redshift in Rindler spacetime). Then the thermal path integral is manifestly equal to the Lorentzian thermal partition function, with the state space identifiable directly as the static field configurations:
\begin{align}
\int_{\phi(\mathbf{x},\tau)=\phi(\mathbf{x},\tau+\beta)} \left[\mathcal{D}\phi(\mathbf{x},\tau)\right] e^{-\int_{0}^{\beta} dt \int d\mathbf{x}\mathcal{L}} \qquad \to \qquad \int \left[\mathcal{D}\phi(\mathbf{x})\right] e^{- \beta V} \equiv \Tr e^{-\beta H},
\end{align}
where only the potential energy $V$ remains in the static case.
\\~\\
As an instructive example of the canonical treatment of higher-derivative theories, consider the higher-derivative Lagrangian of \eqref{mwphi}:
\begin{equation}
\mathcal{L} = \frac{1}{2}\dot{\phi} s\Delta \dot{\phi}.
\end{equation}
Its equations of motion are given by
\begin{equation}
\delta \phi \quad \Rightarrow \quad \partial_\tau \left(s\Delta \partial_\tau \phi\right) = 0,
\end{equation}
which can be solved in general as
\begin{equation}
\label{solri}
\phi = \frac{1}{s\Delta}f(\mathbf{x})t + g(\mathbf{x}),
\end{equation}
using the uniqueness of solutions of elliptic differential equations, and introducing two arbitrary spatial functions $f$ and $g$. Quantization is done by elevating these functions to quantum operators. This expression is a generalization of the 2d case, where $\dot{\phi}$ is just the charge. We will see that it holds here as well. \\
As the Lagrangian is only second-order in time-derivatives, we can write:
\begin{align}
\pi_\phi &= s\Delta \dot{\phi} = f(\mathbf{x}),\\
\mathcal{H} &= \frac{1}{2} \dot{\phi} s\Delta \dot{\phi} = \frac{1}{2} f(\mathbf{x}) \frac{1}{s\Delta} f(\mathbf{x}).
\end{align}
Fourier-expanding $f(\mathbf{x})$, we immediately match with \eqref{thpiri}, identifying $f(\mathbf{x})\equiv \mq(\mathbf{x})$ with the spatial charge distribution on the horizon.
The equal-time CCR boil down to
\begin{equation}
\left[\phi(\mathbf{x},t),\pi_\phi(\mathbf{y},t)\right] = \left[g(\mathbf{x}),f(\mathbf{y})\right] = i\delta(\mathbf{x}-\mathbf{y}).
\end{equation}
Time-evolution of $\phi$ is generated by $H$ as
\begin{equation}
\delta \phi = i\epsilon \left[\int d\mathbf{y} \, \mathcal{H}(\mathbf{y}), \phi(\mathbf{x},0)\right] = \epsilon\int d\mathbf{y} \, \delta(\mathbf{x}-\mathbf{y}) \frac{1}{s\Delta} f(\mathbf{y}) = \epsilon \frac{1}{s\Delta} f(\mathbf{y}),
\end{equation}
which indeed matches with \eqref{solri}.

\subsection{Flat Space}
\label{s:flkernel}
Let us now specify to a planar boundary surface in flat space as in section \ref{s:warmup}. In this case, the kernel is not time-independent, and we have to resort to a more complicated analysis. However, the CCR between $\phi$ and $\mj^0$ together with the equations of motion is enough to determine the structure of the Hilbert space. \\
The classical phase space can be identified as the set of all initial conditions for the equations of motion, or the set of all integration constants. Acting with $\partial_\alpha$ on \eqref{eom2}, we get:
\begin{equation}
\Box \phi + \int dy \partial_{\alpha x}G^{\alpha\beta}(y-x) J^{\beta} = 0,
\end{equation}
using translation invariance in flat space to write $G(y,x) = G(y-x)$. Changing the derivative into one that acts on $y$ instead, integrating by parts, and using $\eqref{eom1}$, one finds $\Box \phi =0$. So the field $\phi$ satisfies the massless Klein-Gordon equation $\Box \phi = 0$ and can be expanded in the standard normal mode expansion on the boundary. The fields $\mj^i$ do not introduce new integration constants, as they are fully determined from those of $\phi$ by \eqref{eom2}: they arise as constraint equations.
\\~\\
The resulting Fourier modes $\phi_k$ can be used to construct the Hilbert space, in combination with its conjugate $\mj^0_k$. The equal-time commutation relation implies
\begin{equation}
\left[\phi(\mathbf{x},t),\mj^0(\mathbf{y},t)\right] = i\delta(\mathbf{x}-\mathbf{y}) \quad \Rightarrow \quad \left[\phi_k,\mj^0_{-k'}\right] = i\delta_{kk'}.
\end{equation}
The modes are linked by \eqref{eom2} as
\begin{equation}
k^0 k_n \phi_k = \mj^0_k,
\end{equation}
satisfying the algebras $\left[\mj^0_k, \mj^0_{-k'}\right] = i k_0k_n\delta_{kk'}$ and $\left[\phi_k, \phi_{-k'}\right] = \frac{i}{k_0k_n}\delta_{kk'}$. Raising and lowering operators are obtained by the Hermiticity requirement:
\begin{equation}
\phi^\dagger_k = \phi_{-k}.
\end{equation}
A much more convenient normalization of the oscillators is found by setting
\begin{equation}
\tilde{\phi}_k = \sqrt{k_0k_n}\phi_k, \quad \tilde{\mj}^0_k = \frac{1}{\sqrt{k_0k_n}} \mj^0_k,
\end{equation}
which satisfy standard commutation relations and are equal: $\phi_k = \mj^0_k$.
\\~\\
As before, the Hilbert space can be seen as $\left\{\left|\phi(\mathbf{x})\right\rangle\right\}$ or $\left\{\left|\mj^0(\mathbf{x})\right\rangle\right\}$. \\
This picture is in agreement with the conclusion of Donnelly and Freidel that only this set of variables is added. In particular, spatial currents $\mj^i$ are \emph{not} canonical variables. In our framework, we explain this due to their role as constraints instead of dynamical degrees of freedom.

\section{Maxwell in Rindler}
\label{app:maxrind}
The purpose of this appendix is to derive formulas \eqref{jat} and \eqref{jai} i.e. to find the linear relation $A[\mj]_\alpha$ for the specific case of Maxwell theory in Rindler space with the horizon as boundary. In Lorentz gauge 
\begin{equation}
\nabla^\mu A_\mu=0,
\end{equation}
the Maxwell equations of motion $\nabla^\mu F_{\mu\nu}=0$ reduce to
\begin{equation}
\nabla^\mu \nabla_\mu A_\nu=0.\label{mweom}
\end{equation}
The Rindler coordinate system in $d$ dimensions is the metric:
\begin{equation}
ds^2=-\rho^2 dt^2+d\rho^2+{d\mathbf{x}}^2 = e^{2r}(-dt^2+dr^2),\label{rindmetric}
\end{equation}
where $\rho = e^{r}$ and $\mathbf{x}=\{x^i\}$ denotes all coordinates parallel to the horizon i.e. the $d-2$ spectator dimensions. The only non-vanishing Christoffel symbols are $\Gamma^{t}_{t\rho} = 1/\rho$ and $\Gamma^{\rho}_{tt}=-\rho$. The boundary of the theory is placed close to the horizon at $\rho=\epsilon \to 0$ or $r=r_* \to -\infty$. Upon Wick rotating this becomes the tensor product of flat space in polar coordinates with the $d-2$ spectator dimensions:
\begin{equation}
ds^2=\rho^2d\tau^2+d\rho^2+{d\mathbf{x}}^2.
\end{equation}
Introducing elementary scalar modes:
\begin{equation}
\label{elmode}
\phi_{\omega, \mathbf{k}}=\frac{\sqrt{\sinh(\pi\omega)}}{(2\pi)^{\frac{D-2}{2}}\pi}K_{i\omega}(k\rho)e^{i\mathbf{k}\cdot \mathbf{x}}e^{-i\omega t},
\end{equation}
solving $\Box \phi_{\omega,\mathbf{k}}= \frac{1}{\rho^2}\left(-\partial_t^2+ (\rho\partial_\rho)^2- k^2\rho^2\right)\phi = 0$ with $k = \left|\mathbf{k}\right|$, and introducing a basis of unit vectors along the $d-2$ trivial directions as:
\begin{equation}
\e^{(k)}_\mu = \left(0,0,\mathbf{n}^{(k)}\right), \quad\, \e^{(a)}_\mu = \left(0,0,\mathbf{n}^{(a)}\right),
\end{equation}
with $\mathbf{n}^{(a)}$ short for an orthonormal $d-3$ dimensional basis orthogonal to $\mathbf{n}^{(k)}=\mathbf{k}/k$, one can solve the bulk equations of motion $\nabla^\mu\nabla_\mu A_\nu=0$ by the complete set of modes: 
\begin{gather}
\begin{aligned}
A_{\mu,\omega\mathbf{k}}^{(1)} &= \frac{1}{k}\left(\rho\partial_\rho, \frac{1}{\rho}\partial_t, \mathbf{0}\right)\phi_{\omega, \mathbf{k}}\\
A_{\mu,\omega\mathbf{k}}^{(0)} &= \frac{1}{k}\partial_\mu\phi_{\omega, \mathbf{k}}\\
A_{\mu,\omega\mathbf{k}}^{(a)} &= \e^{(a)}_\mu \phi_{\omega, \mathbf{k}},
\end{aligned}\label{5.5}
\end{gather}
and
\begin{equation}
A_{\mu,\omega\mathbf{k}}^{(k)}= \e^{(k)}_\mu \phi_{\omega, \mathbf{k}}.
\end{equation}
This last solution though, does not satisfy Lorenz gauge \eqref{mwgauge} and is thus not to be considered. 

The residual gauge freedom of Maxwell theory in Lorenz gauge is $A_\mu\sim A_\mu +\partial_\mu \phi$, with $\phi$ satisfying $\Box \phi=0$. This residual gauge freedom is precisely captured by the modes $A^{(0)}_{\omega\mathbf{k}}$. To obtain an isomorphism between $A$ and $\mj$ one has to completely gauge-fix $A$ i.e. choose one representative in each gauge orbit. We will take the most natural choice to construct the field $A$ out of only the modes $A^{(1)}_{\omega\mathbf{k}}$ and $A^{(a)}_{\omega\mathbf{k}}$, so we turn off the modes $A^{(0)}_{\omega\mathbf{k}}$.

The next step is to split the bulk field from the edge field. The bulk photon obeys PMC boundary conditions $n_\mu F^{\mu\nu} = 0$. This constrains the allowed range of $\omega$ and $\mathbf{k}$ in the solution space \eqref{5.5}. More in particular this constrains the field $\phi$ to be Dirichlet $\phi\rvert_\text{bdy}=0$ in the expression for the modes $A^{(1)}_{\omega\mathbf{k}}$ and Neumann $\rho\partial_\rho \phi\rvert_\text{bdy}=0$ for the modes $A^{(a)}_{\omega\mathbf{k}}$. The result is the bulk photon with $d-2$ polarizations that satisfies PMC boundary conditions.
\\~\\
Now for the interesting part. The edge field is obtained by finding a particular solution that satisfies the boundary conditions \eqref{mwbc}:
\begin{equation}
\left.\left(\sqrt{-g}n_\mu F^{\mu\nu}\right)\right|_\text{bdy}=\mj^\nu.
\end{equation}
Specified to Rindler these read\footnote{The vector $n$ is the outwards normal to $\mathcal{M}$ and hence points to decreasing values of $\rho$.}
\begin{equation}
\left.-g^{\alpha\alpha}\rho \left(\partial_\rho A_\alpha-\partial_\alpha A_\rho\right)\right|_\text{bdy}=\mj^\alpha.\label{mwbcA}
\end{equation}
Since $\mj$ is necessarily static (it lives on the boundary), the same goes for the edge field $A[\mj]$ isomorphic to it. It is useful to decompose $\mj$ into Fourier components and polarizations as:
\begin{equation}
\mj =\sum_\mathbf{k}\left(\mj^{(k)}_\mathbf{k}\e^{(k)}+\mj^{(a)}_\mathbf{k}\e^{(a)}+\mq_\mathbf{k}\e^{(0)}\right)e^{i\mathbf{k}\cdot \mathbf{x}},\label{mwjexpansion}
\end{equation}
where we introduced the notation $\mj^t=\mq=\sum_\mathbf{k}\mq_\mathbf{k}e^{i\mathbf{k}\cdot \mathbf{x}}$ and $\e^{(0)}_\mu=g^{tt}(1,0,\mathbf{0})$. Notice that for $\mj$ to be an acceptable source for Maxwell theory; it must be conserved: $\partial_\alpha \mj^\alpha$=0. For the static current \eqref{mwjexpansion} this becomes just $\partial_i \mathcal{J}^i=0$. This enforces through \eqref{mwjexpansion} $\mj^{(k)}_\mathbf{k}=0$, and indeed clearly no $k$ component of $\mj$ can be created using the bulk solutions $A^{(1)}_\mathbf{k}$ and $A^{(a)}_\mathbf{k}$. We introduced here the convention that when the $\omega$ subscript is dropped, the zero modes $\omega=0$ of \eqref{5.5} are implied. More in particular the zero mode sector of Maxwell theory is 
\begin{equation}
A=\sum_\mathbf{k}\left(c^{(a)}_\mathbf{k}A^{(a)}_\mathbf{k}+c^{(1)}_\mathbf{k}A^{(1)}_\mathbf{k}\right)e^{i\mathbf{k}\cdot\mathbf{x}},\label{mwAexpansion}
\end{equation}
with 
\begin{equation}
A^{(a)}_{\mu,\mathbf{k}}=\e^{(a)}_\mu\phi_k,\quad A^{(1)}_{\mu,\mathbf{k}}=\left(\rho\partial_\rho,0,\mathbf{0}\right)\phi_k
\end{equation}
the zero modes of \eqref{5.5} and $\phi_k$ the $\omega=0$ solution of $\Box \left(\phi_ke^{i\mathbf{k}\cdot\mathbf{x}}\right)=0$ which reduces to $(\rho\partial_\rho)^2\phi_k=\rho^2k^2\phi_k=-\rho^2\Delta \phi_k$. We conveniently choose the normalization such that ${\phi_k}\rvert_\text{bdy}=1$:\footnote{This is a choice, with no influence on what follows.}
\begin{equation}
\phi_k=\frac{K_0(k\rho)}{K_0(k\epsilon)}.
\end{equation}
For future reference we introduce a notation for the normal derivative of $\phi_k$ at the boundary:
\begin{equation}
\left.{\rho\partial_\rho\phi_k}\right|_\text{bdy}=-K_0^{-1}(k\epsilon)=\ln^{-1}\frac{k\epsilon}{2} \to {r_*}^{-1}=-s^{-1},\label{mws}
\end{equation}
where $r_*$ is the regulator for the location of the boundary in tortoise coordinates and where we used the small argument expansion of the Bessel function $K_0$. The limit $\epsilon\to 0$ can be equivalently enforced as $s\to \infty$. 

The task at hand has been reduced to finding a relation between the expansion coefficients $c$ in \eqref{mwAexpansion} and the expansion coefficients of the current in \eqref{mwjexpansion}. The $t$ component of the boundary condition \eqref{mwbcA} reduces to $\left.-\rho \partial_\rho A_t\right|_\text{bdy} = \mj_t = \left.g_{tt}\right|_\text{bdy} \mq$, or inserting the mode expansions:
\begin{equation}
\left.c^{(1)}_\mathbf{k}(\rho\partial_\rho)^2 \phi_k \right|_\text{bdy}=-\left.g_{tt}\right|_\text{bdy} \mq_\mathbf{k}.
\end{equation}
Using the equations of motion and normalization of $\phi_k$, and the Rindler metric \eqref{rindmetric} one obtains the desired relation:
\begin{equation}
c^{(1)}_\mathbf{k}=\frac{1}{k^2}\mq_\mathbf{k}. 
\end{equation}
In terms of the full field component $A_t$ this becomes
\begin{equation}
A[\mj]_t = \sum_\mathbf{k}\frac{1}{k^2}\mq_\mathbf{k}\rho\partial_\rho\phi_k e^{i\mathbf{k}\cdot \mathbf{x}}.\label{mwPO1}
\end{equation}
Evaluation at the boundary using \eqref{mws} results in 
\begin{equation}
\left.\mj^tA[\mj]_t\right|_\text{bdy}=\mq\frac{1}{s\Delta}\mq,\label{mw1}
\end{equation}
which is \eqref{jat} in the main body.

A similar analysis relates $c^{(a)}_\mathbf{k}$ to $\mj^{(a)}_\mathbf{k}$. The spatial part of the boundary condition \eqref{mwbcA} is $\left.-\rho\partial_\rho A_i\right|_\text{bdy}=\mj_i=\mj^i$, where we already used the triviality of the Rindler metric for the spectator dimensions. We obtain
\begin{equation}
\left.-c^{(a)}_\mathbf{k}\rho\partial_\rho\phi_k\right|_\text{bdy}=\mj^{(a)}_\mathbf{k},
\end{equation}
or using \eqref{mws}:
\begin{equation}
c^{(a)}_\mathbf{k}=s\mj^{(a)}_\mathbf{k}.
\end{equation}
In terms of the spatial field components $A_i$ this becomes
\begin{equation}
A[\mj]_i=\sum_\mathbf{k}s\mj^{(a)}_\mathbf{k}\e^{(a)}_i\phi_ke^{i\mathbf{k}\cdot\mathbf{x}}. \label{mwPO2}
\end{equation}
Evaluation at the boundary results in
\begin{equation}
\left.\mj^iA[\mj]_i\right|_\text{bdy}=s\mathcal{J}^i \mathcal{J}^i,\label{mw2}
\end{equation}
which is \eqref{jai} in the main body.

\section{Evaluation of Path Integral}
\label{app:AS}
The evaluation of the following path integral
\begin{equation}
\label{eqpi}
Z = \int \left[\dpi \mq\right]\left[\dpi g\right]\exp{-\int_0^\beta d\tau \Tr(\frac{a}{2}\mq^2+i\mq \partial_\tau g g^{-1})}
\end{equation}
was largely done by Alekseev, Faddeev and Shatashvili in \cite{Alekseev:1990mp,Alekseev:1988vx}, which we review here. The main difference is our choice of Hamiltonian as the Casimir, instead of a Cartan element. We focus on $SU(n)$, with $g(\tau) \in SU(n)$ and $\mq(\tau) \in \mathfrak{su}(n)$ and the trace in the defining representation. \\

Diagonalizing the matrix $\mq = f \mq_0 f^{-1}$ into a diagonal matrix $\mq_0$ and the basis of eigenvectors $f$, and writing the path-integral measure as $\left[\dpi \mq\right] \to \left[\dpi \mq_0\right]\left[\dpi f\right]$, one can write the above path-integral \eqref{eqpi} as (redefining $g \to f g$):
\begin{equation}
Z = \int \left[\dpi \mq_0\right]\left[\dpi f\right]\left[\dpi g\right]\exp{-\int_0^\beta d\tau \Tr(\frac{a}{2}\mq_0^2+i\mq_0 \partial_\tau g g^{-1} + i\mq_0 f^{-1}\partial_\tau f)}.
\end{equation}
Denoting the diagonal elements of $Q_0$ as $m_i^0$, ordered as $m_1^0 \geq m_2^0 \geq  ... \geq m^0_n$, and using the decomposition \cite{Alekseev:1990mp,Alekseev:1988vx}
\begin{equation}
\Tr \mq_0 \partial_\tau g g^{-1} = \sum_{i=1}^{n}m_i^0 d\phi_i^0 + \sum_{i,k}m_i^k d\phi_i^k, \qquad \Tr \mq_0 f^{-1}\partial_\tau f = \sum_{i,k}n_i^k d\bar\phi_i^k,
\end{equation}
in terms of $2\pi$-periodic angular variables $\phi_i^0$, $\phi_i^k$ and $\bar{\phi}_i^k$, and with $m_i^{k-1} \geq m_i^k \geq m_{i+1}^{k-1}$ and $n_i^{k-1} \geq n_i^k \geq n_{i+1}^{k-1}$, we find
\begin{align}
\label{pipq}
Z = \int [\dpi \Delta_i]&[\dpi \phi_i^0][\dpi \Delta_i^k][\dpi \phi_i^k][\dpi \bar{\Delta}_i^k][\dpi \bar{\phi}_i^0] \nonumber \\
&\times \exp{-\int_0^\beta d\tau \sum_m \frac{a}{2}(m_i^0)^2 + i \sum_{i=1}^{r}\Delta_i d\phi_i^{0} + i \sum_{i,k}\Delta_i^k d\phi_i^{k} + i \sum_{i,k}\bar{\Delta}_i^k d\bar{\phi}_i^{k}},
\end{align}
where $\Delta_i = m_i^0-m_n^0$, $\Delta_i^k = m_i^k - m_n^0$ and $\bar{\Delta}_i^k = n_i^k + m_n^0$; this shift arising from the constraint of tracelessness of $\mathfrak{su}(N)$. 
\\~\\
The expression \eqref{pipq} is interpretable as a phase space path integral of multiple free particles on independent circles with coordinates $\phi_i^{0}, \phi_i^{k}, \bar{\phi}_i^{k}$ ($\phi \sim \phi + 2\pi$), generalizing this interpretation from $U(1)$ where $\oint A$ and $E$ are conjugate variables that, upon charge quantization, are phase-space coordinates on a circle (see e.g. \cite{Donnelly:2012st}). As a result, the ``momenta'' $\Delta_i, \Delta_i^k, \bar{\Delta}_i^k$ are quantized.
\\~\\
The path integral over the $\phi^0_i$ gives a sum over an $r$-dimensional (non-negative) integer-valued vector with components $\Delta_i$ that labels the lowest-weight state underlying the representation: it is the first row of the Gelfand-Tsetlin table. The Dynkin labels are then found as $\lambda_i = \Delta_i-\Delta_{i-1}$. Path-integrating over the $\phi_i^k$- and $\bar{\phi}_i^k$-variables, yields a (non-negative) integer-valued distribution of $\Delta_i^k$'s and $\bar{\Delta}_i^k$'s forming two separate Gelfand-Tsetlin tables, that count all weights in the representation. \\
The first term in the action of \eqref{pipq} depends only on the lowest-weight parameters $m_i^0$ and evaluates to the Casimir of the irrep, in the end giving
\begin{equation}
Z=\sum_R (\dim R)^2e^{-A C_{R}}.
\end{equation}
One of the $\text{dim }R$ factors arises from the $g$-path integral, the other roughly from the off-diagonal elements of $\mq$ (the eigenvector basis $f$). Taking e.g. a fixed diagonal $\mq$ results in the coadjoint orbit action of the element $\mq$, and the path integral evaluates to a single character of the irrep labeled by the diagonal matrix $\mq$ \cite{Alekseev:1990mp}.
\\~\\
Take as an example $SU(2)$, where only one Cartan element exists. Due to tracelessness, the $2\times 2$ $\mq_0$-matrix eigenvalues are then $+b$ and $-b$ for some real number $b$. $\Delta$ is a non-negative integer, requiring $2b = 0,1,2...$, so we can set $b = j$ for $j=0,1/2,1,...$. The Dynkin label of the irrep is $\lambda = \Delta = 2j$. The Gelfand-Tsetlin table is then 
\begin{align}
&\quad 2j \quad 0 \nonumber \\
&\qquad \Delta^1 
\end{align}
for $\Delta^1 = 0 \hdots 2j$, forming a $2j+1$-dimensional representation. The first term in the action of \eqref{pipq} is proportional to $b^2 = j^2$, which is the classical value of the Casimir. This is different from the (quantum) Casimir: $C_j = j^2+j$. Classically, one can set $\mathbf{J}_x=\mathbf{J}_y=0$ and $\mathbf{J}_z=\pm j$ with hence $\mathbf{J}^2 = \mathbf{J}_x^2 + \mathbf{J}_y^2 + \mathbf{J}_z^2=j^2$, which is of course impossible quantum-mechanically. The mismatch can be understood as a quantum renormalization effect, arising from the path integral measure as a regularization artifact of functional determinants, that should be taken into account. This is quite standard when evaluating coadjoint orbit path integrals \cite{Oblak:2016eij,Gukov:2008ve}: it is the famous Weyl shift problem $\lambda \to \lambda + \rho$, with $\rho$ the Weyl vector.

\section{Classical Solution of 2d Yang-Mills on a Disk}
\label{app:class}
We provide details on the statement that the field $A_t^a=-\frac{\rho^2}{2}\mq^a$ and $A_\rho^a=0$ solves the 2d Yang-Mills equations of motion $(D^\mu F_{\mu\nu})^a=0$ and satisfies Lorenz gauge $(D^\mu A_\mu)^a=0$. By construction, it satisfies the correct boundary condition.

Writing out the derivative explicitly in terms of the Christoffel and Yang-Mills connections, one obtains the equations of motion:
\begin{equation}
g^{\mu\alpha}\left(\partial_\alpha F_{\mu\nu}^a-\Gamma^\beta_{\alpha\mu}F_{\beta\nu}^a-\Gamma^\beta_{\alpha\nu}F_{\mu\beta}^a\right)+\tensor{f}{^a_{bc}}A^{\mu b}F_{\mu\nu}^c=0.
\end{equation}
Inserting the proposed particular solution, and taking the component $\nu=t$, the $\tensor{f}{^a_{bc}}$-part drops out. The Rindler Christoffel symbols are $\Gamma^t_{t\rho}=1/\rho$ and $\Gamma^\rho_{tt}=-\rho$ with all others vanishing. One arrives at 
\begin{equation}
\partial_\rho F_{\rho t}^a-\frac{1}{\rho} F_{\rho t}^a=-\mq^a+\mq^a=0.
\end{equation}

The particular solution satisfies $\nabla_t F_{t\rho}^a=0$ and $\nabla_\rho F_{\rho\rho}^a=0$ since it is static, and because $F$ is anti-symmetric. The $\nu=\rho$ component of the equations of motion hence reduces to only the $\tensor{f}{^a_{bc}}$-part, which reads:
\begin{equation}
\frac{\rho}{2}\tensor{f}{^a_{bc}}\mq^b\mq^c=0,
\end{equation}
since $\tensor{f}{^a_{bc}}$ is also anti-symmetric. \\

Using $\tensor{f}{^a_{bc}}g^{\mu\nu}A_\mu^b A_\nu^c=0$, and writing out the Lorenz gauge constraint by inserting the particular solution, we obtain:
\begin{equation}
\nabla^\mu A_\mu^a+\tensor{f}{^a_{bc}}g^{\mu\nu}A_\mu^b A_\nu^c=\nabla^\mu A_\mu^a=0,
\end{equation}
where the last equality is trivial: the particular solution reduces to just the Maxwell particular solution which we know to be divergence-free with respect to only the Christoffel connection. This completes the proof.

\section{Boundary Correlators in 2d Maxwell}
\label{sect:u1}
As a specific application of the discussion on boundary-anchored Wilson lines in Yang-Mills, it is instructive to return to the Maxwell case. As all integrals are Gaussian, both in the Maxwell bulk as on the boundary particle-on-U(1) theory, we can calculate the boundary correlators directly without resorting to a dimensional reduction of 2d WZW correlators. The relevant bilocal operators on the boundary of 2d Maxwell are:
\begin{equation}
\mathcal{O}_q(\tau_i,\tau_f)=\exp{i q\phi(\tau_f)-i q\phi(\tau_i)}=\exp{i q\int_{\tau_i}^{\tau_f}d\tau \partial_\tau \phi} \equiv \mathcal{W}_q(\tau_i,\tau_f).
\end{equation}
As an illustration, we explicitly match the bulk and boundary calculation of both the two-point function and the crossed four-point function.
\\~\\
Consider first the two-point function. The bulk calculation follows from the diagram:
\begin{align}
\begin{tikzpicture}[scale=1, baseline={([yshift=0cm]current bounding box.center)}]
\draw[thick] (0,0) circle (1.5);
\draw[thick,blue] (-1.3,-.7) arc (120:60:2.6);
\draw[fill,black] (-1.3,-.68) circle (0.1);
\draw[fill,black] (1.3,-.68) circle (0.1);
\draw (0,0) node {\footnotesize \color{blue}$q$};
\draw (0,1) node {\footnotesize  $q_2$};
\draw (0,-1) node {\footnotesize  $q_1$};
\draw (-1.75,-.75) node {\footnotesize $\tau_i$};
\draw (1.75,-.75) node {\footnotesize $\tau_f$};
\draw (0,-2) node {$I_1$};
\end{tikzpicture}~~\raisebox{-3pt}{$\ \ \  = \ \ \Big\langle \mathcal{W}_q(\tau_i,\tau_f)\Big\rangle .$}~~~
\end{align}
Explicitly this is:
\begin{align}
\Big\langle \mathcal{W}_q(\tau_i,\tau_f)\Big\rangle &= \int d q_1 d q_2 e^{-A_1 C_{q_1}}e^{-A_2 C_{q_2}}\delta (q_1+q-q_2) \label{mw2pt1stform} \\
&=\int d q_1 e^{-A_1 C_{q_1}}e^{-A_2 C_{q_1+q}}.\label{mw2pt2ndform}
\end{align}
The boundary calculation is just the path integral
\begin{equation}
\Big\langle \mathcal{O}_q(\tau_i,\tau_f) \Big\rangle = \int \left[\dpi \phi \right]\exp{i q\int_{\tau_i}^{\tau_f}d\tau \partial_\tau \phi}\exp{-\frac{1}{2 a}\int d\tau \partial_\tau \phi \partial_\tau \phi}.\label{mw2pt2}
\end{equation}
The inserted bilocal operator acts as a source term in the boundary action extracting a charge $q$ at $\tau_i$, and re-injecting this same charge at $\tau_f$. This Gaussian path integral can be computed directly by solving the sourced equations of motion and inserting the solution into the action: one recovers \eqref{mw2pt2ndform}. \\
It is possible to go from \eqref{mw2pt2} directly to the diagrammatic expression \eqref{mw2pt1stform} without passing through \eqref{mw2pt2ndform}. Splitting the path integral \eqref{mw2pt2}, using
\begin{equation}
\int_{I_1}\left[\dpi q\right]\exp{-\int_{I_1}d\tau \left(\frac{a}{2}q_1^2-iq_1\partial_\tau \phi\right)}=\exp{-\frac{1}{2 a}\int_{I_1} d\tau \partial_\tau \phi \partial_\tau \phi},\label{expand}
\end{equation}
with $I_1$ shorthand for the interval from $\tau_i$ to $\tau_f$ and doing the same for $I_2$, one obtains:
\begin{align}
\Big\langle \mathcal{O}_q(\tau_i,\tau_f) \Big\rangle=\int \left[\dpi q_1\right]\left[\dpi q_2\right]\left[\dpi \phi\right]&\exp{-\int_{I_2}d\tau \left(\frac{a}{2}q_2^2-iq_2\partial_\tau \phi\right)}\nonumber\\&\cross\exp{-\int_{I_1}d\tau \left(\frac{a}{2}q_1^2-i(q+q_1)\partial_\tau \phi\right)}.\label{mw2pt3}
\end{align}
At this point $\phi$ is just a Lagrange multiplier field; its path integral enforces charge conservation. Within the interval $I_i$ this results in $\delta(\partial_\tau q_i)$. On the boundary $\partial I_1$ the $\phi$ path integral results in $\delta(q_1+q-q_2)$. Using the dictionary $A_i=a L_i$ as before completes the proof.
\\~\\
As a second example consider the crossed four point function. The bulk calculation follows from the diagram
\begin{align}
\begin{tikzpicture}[scale=1, baseline={([yshift=0cm]current bounding box.center)}]
\draw[thick]  (0,0) ellipse (1.6 and 1.6);
\draw[thick,blue] (1.1,-1.2) arc (35.7955:82:4);
\draw[thick,blue] (-1.1,-1.2) arc (144.2045:98:4);
\draw (0,-0.8) node {\small $q_3$};
\draw (0,1) node {\small $q_1$};
\draw (-1,-0.2) node {\small $q_2$};
\draw (1,-0.2) node {\small $q_4$};
\draw (-0.5,0.35) node {\small \color{blue}$q_A$};
\draw (0.5,0.35) node {\small \color{blue}$q_B$};
\draw[fill,black] (-1.56,0.4) circle (0.1);
\draw[fill,black] (1.56,0.4) circle (0.1);
\draw[fill,black] (-1.11,-1.17) circle (0.1);
\draw[fill,black] (1.11,-1.17) circle (0.1);
\draw (-1.9,0.4) node {\small $\tau_1$};
\draw (-1.45,-1.4) node {\small $\tau_2$};
\draw (1.45,-1.4) node {\small $\tau_3$};
\draw (1.9,0.4) node {\small $\tau_4$};
\draw (-2.1,-0.6) node {$I_2$};
\draw (2.1,-0.6) node {$I_4$};
\draw (0,-2) node {$I_3$};
\end{tikzpicture} ~~\raisebox{-3pt}{$\ \ \  = \ \ \Big\langle \mathcal{W}_{q_A}(\tau_1,\tau_3){W}_{q_B}(\tau_2,\tau_4)\Big\rangle .$}~~~
\end{align}
Explicitly using the 2d Maxwell diagrammatic rules this is:
\begin{equation}
\Big\langle \mathcal{W}_{q_A}(\tau_1,\tau_3){W}_{q_B}(\tau_2,\tau_4)\Big\rangle=\prod_{i}\left(\int d q_i e^{-A_i C_{q_i}}\right)\delta(q_1+q_A-q_2)\delta(q_2+q_B-q_3)\delta(q_3-q_A-q_4).\label{mw4ptcross1}
\end{equation}
The boundary calculation is the path integral
\begin{align}
\nonumber\Big\langle \mathcal{O}_{q_A}(\tau_1,\tau_3){O}_{q_B}(\tau_2,\tau_4)\Big\rangle= \int \left[\dpi \phi \right]&\exp{-\frac{1}{2 a}\int d\tau \partial_\tau \phi \partial_\tau \phi}\exp{i\int_{I_2}q_A d\tau \partial_\tau \phi}\\\nonumber&\cross\exp{i\int_{I_3}(q_A+q_B) d\tau \partial_\tau \phi}\exp{i\int_{I_4}q_B d\tau \partial_\tau \phi}.
\end{align}
Splitting up this path integral using \eqref{expand}, one obtains the analogue of \eqref{mw2pt3}. Path integration over the Lagrange multiplier $\phi$ enforces current conservation, resulting in the four ordinary integrals over the $q_i$'s and the three delta functions of \eqref{mw4ptcross1}. The dictionary $A_i=a L_i$ does the rest.

\end{document}